\documentclass[a4paper,fleqn,usenatbib]{mnras}
\usepackage{Times}
\usepackage[T1]{fontenc}
\usepackage{ae,aecompl}

\usepackage{graphicx}	
\usepackage{amsmath}	
\usepackage{amssymb}	

\title[Evolution of LAEs with SC4K]{The evolution of rest-frame UV properties, Ly$\alpha$ EWs and the SFR-Stellar mass relation at $\bf z\sim2-6$ for SC4K LAEs}

\author[Santos et al.]{S. Santos$^{1}$\thanks{E-mail: s.santos@lancaster.ac.uk}, D. Sobral$^{1}$, J. Matthee$^{2}$\thanks{Zwicky Fellow}, J. Calhau$^{1}$, E. da Cunha$^{3,4,5}$, B. Ribeiro$^{6}$, \newauthor A. Paulino-Afonso$^{7}$, P. Arrabal Haro$^{8,9}$, J. Butterworth$^{1}$\\ 
$^{1}$ Department of Physics, Lancaster University, Lancaster, LA1 4YB, UK \\
$^{2}$ Department of Physics, ETH Z$\ddot{u}$rich,Wolfgang-Pauli-Strasse 27, 8093 Z$\ddot{u}$rich, Switzerland\\
$^{3}$ International Centre for Radio Astronomy Research, University of Western Australia, 35 Stirling Hwy, Crawley, WA 6009, Australia \\
$^{4}$ Research School of Astronomy and Astrophysics, The Australian National University, Canberra, ACT 2611, Australia \\
$^{5}$ ARC Centre of Excellence for All Sky Astrophysics in 3 Dimensions (ASTRO 3D) \\
$^{6}$ Leiden Observatory, Leiden University, P.O. Box 9513, NL-2300 RA Leiden, The Netherlands\\
$^{7}$ CENTRA - Centro de Astrof\'{i}sica e Gravita\c{c}\~{a}o, Instituto Superior T\'{e}cnico, Av. Rovisco Pais, 1, 1049-001, Lisboa, Portugal \\
$^{8}$ Instituto de Astrof\'{i}sica de Canarias (IAC), E-38205 La Laguna, Spain\\
$^{9}$ Departamento de Astrof\'{i}sica, Universidad de La Laguna, E-38206 La Laguna, Spain}

\pubyear{2019}

\begin{document}
\label{firstpage}
\pagerange{\pageref{firstpage}--\pageref{lastpage}}
\maketitle

\begin{abstract}
We explore deep rest-frame UV to FIR data in the COSMOS field to measure the individual spectral energy distributions (SED) of the $\sim4000$ SC4K \citep{Sobral2018} Lyman-$\alpha$ (Ly$\alpha$) emitters (LAEs) at $z\sim2-6$. We find typical stellar masses of 10$^{9.3\pm0.6}$\,M$_{\odot}$ and star formation rates (SFR) of SFR$_{\rm SED}=4.4^{+10.5}_{-2.4}$\,M$_{\odot}$\,yr$^{-1}$ and SFR$_{\rm Ly\alpha}=5.9^{+6.3}_{-2.6}$\,M$_{\odot}$\,yr$^{-1}$, combined with very blue UV slopes of $\beta=-2.1^{+0.5}_{-0.4}$, but with significant variations within the population. M$_{\rm UV}$ and $\beta$ are correlated in a similar way to UV-selected sources, but LAEs are consistently bluer. This suggests that LAEs are the youngest and/or most dust-poor subset of the UV-selected population. We also study the Ly$\alpha$ rest-frame equivalent width (EW$_0$) and find 45 ``extreme" LAEs with EW$_0>240$\,{\AA} (3\,$\sigma$), implying a low number density of $(7\pm1)\times10^{-7}$\,Mpc$^{-3}$. Overall, we measure little to no evolution of the Ly$\alpha$ EW$_0$ and scale length parameter ($w_0$) which are consistently high (EW$_0=140^{+280}_{-70}$\,\AA, $w_0=129^{+11}_{-11}$\,\AA) from $z\sim6$ to $z\sim2$ and below. However, $w_0$ is anti-correlated with M$_{\rm UV}$ and stellar mass. Our results imply that sources selected as LAEs have a high Ly$\alpha$ escape fraction (f$_{\rm esc, Ly\alpha}$) irrespective of cosmic time, but f$_{\rm esc, Ly\alpha}$ is still higher for UV-fainter and lower mass LAEs. The least massive LAEs ($<10^{9.5}$ M$_{\odot}$) are typically located above the star formation ``Main Sequence" (MS), but the offset from the MS decreases towards $z\sim6$ and towards $10^{10}$\,M$_{\odot}$. Our results imply a lack of evolution in the properties of LAEs across time  and reveals the increasing overlap in properties of LAEs and UV-continuum selected galaxies as typical star-forming galaxies at high redshift effectively become LAEs.
\end{abstract}


\section{Introduction} \label{sec:introduction}

The Lyman-$\alpha$ (Ly$\alpha$, $\lambda_{\rm 0,vacuum}=1215.67$\,\AA) emission line has been predicted to be associated with young star-forming galaxies \citep[SFGs, e.g.][]{PartridgePeebles1967} but it can also be emitted by active galaxy nuclei \citep[AGN; e.g.][]{Miley2008,Sobral2018Spectra}. Typical Ly$\alpha$ emitters (LAEs) selected with deep surveys have been found to have low stellar mass (M$_\star\lesssim10^{9}$\,M$_\odot$), low dust content and high specific star formation rates \citep[e.g.][]{Gawiser2006,Gawiser2007}, but LAEs can span a wide range in different properties \citep[e.g.][]{Hagen2016,Matthee2016}. Observationally, the transition between the dominant powering source in LAEs seems to occur at $\sim10^{43}$\,erg\,s$^{-1}$, roughly two times the characteristic Ly$\alpha$ luminosity (L$^\star_{\rm Ly\alpha}$) at $z\sim2-3$ \citep[see][]{Sobral2018Spectra}.

Searches using the Ly$\alpha$ emission line have been extremely successful at selecting young SFGs through narrow band searches \citep[e.g.][]{Hu2004,Ouchi2008,Matthee2015,Santos2016,Sobral2017,Harikane2018,Arrabal2018} and spectroscopically confirming bright LAEs \citep[e.g.][]{Sobral2015,Hu2016,Matthee2017spectra,Sobral2018Spectra,Shibuya2018} due to the bright Ly$\alpha$ feature. Other studies have successfully selected samples of LAEs using integral field spectroscopy observations \citep[e.g.][]{vanBreukelen2005,Blanc2011,Bacon2015,Drake2017} and blind spectroscopy \citep[e.g.][]{Martin2004,Rauch2008,Cassata2011,LeFevre2015}. LAEs typically have faint continua, and thus the study of properties of individual sources has typically only been done for extreme LAEs with L$\gtrsim$L$^\star_{\rm Ly\alpha}$ \citep[e.g.][]{Ouchi2013,Sobral2015}. For $\lesssim$L$^\star_{\rm Ly\alpha}$ LAEs, studies have typically resorted to stacking of sources \citep[e.g.][]{Momose2014,Kusakabe2018}. More commonly, large samples of high-redshift SFGs have been selected by searching for the presence of a Lyman Break \citep[e.g.][]{Steidel1996a,Steidel1999,Madau1996}. Currently, there are $>10,000$s of known galaxies at $z\sim2-10$ \citep[see e.g.][]{Bouwens2014b,Bouwens2015}, mostly consisting of faint sub-L$^\star_{\rm UV}$ galaxies found through deep small area searches, typically too faint to follow-up with current spectroscopic instrumentation.

While Ly$\alpha$ surveys are efficient at selecting galaxies, inferring intrinsic properties of a galaxy directly from its Ly$\alpha$ emission is challenging due to the complex nature of Ly$\alpha$ radiative transfer. Ly$\alpha$ photons suffer resonant scattering from gas in the Interstellar/Circumgalactic Medium (ISM/CGM) and get easily absorbed by dust \citep[for a review on the process of Ly$\alpha$ radiative transfer see][]{Dijkstra2017} which can suppress Ly$\alpha$ emission even in young SFGs. The complex physics of Ly$\alpha$ radiative transfer means that the Ly$\alpha$ escape fraction (f$_{\rm esc, Ly\alpha}$ - the ratio between observed and intrinsic Ly$\alpha$ luminosity) is difficult to predict. Multiple studies have taken different approaches to this problem. Observationally, f$_{\rm esc, Ly\alpha}$ has been measured by comparing Ly$\alpha$ to dust-corrected H$\alpha$ luminosities \citep{Oteo2015,Matthee2016,Sobral2017}. Some studies estimate f$_{\rm esc, Ly\alpha}$ by computing the ratio between star formation rate (SFR) derived from Ly$\alpha$ (assuming case B recombination) and SFR derived from alternative methods such as from spectral energy distributions \citep[SEDs, ][]{Cassata2015} or the far-infrared \citep[FIR, ][]{Wardlow2014}. Others measure the ratio between the observed Ly$\alpha$ luminosity density and the dust-corrected H$\alpha$ luminosity density \citep{Sobral2017}. Alternatively, studies have measured the ratio between Ly$\alpha$ SFR density (SFRD) and UV SFRD by integrating the respective luminosity functions \citep{Sobral2018}. Typical SFGs at $z\sim2-3$ are found to have very low f$_{\rm esc, Ly\alpha}$ \citep[$<5\%$, e.g.][]{Hayes2010,Oteo2015,Cassata2015,Matthee2016}. However, sources selected due to their Ly$\alpha$ emission have much higher f$_{\rm esc, Ly\alpha}$ \citep[as high as $\sim40\%$ at $z=2.2$,][]{Sobral2017}.

Despite the complexity of the Ly$\alpha$ radiative transfer, properties of the Ly$\alpha$ line such as its equivalent width (EW) have been shown to hold important information. Sources selected by their Ly$\alpha$ emission typically have high EWs, with rest-frame Ly$\alpha$ EW (EW$_0$) $\sim50-150$\,\AA\, at $z\sim0.3-6$ \citep[see e.g.][]{Gronwall2007, Hashimoto2017,Wold2017} which can be explained by young stellar ages, low metallicities and/or top-heavy initial mass functions \citep{Schaerer2003,Raiter2010} or complex radiative transfer effects \citep{Neufeld1991}. The high Ly$\alpha$ EW$_0$ measured for LAEs even at low redshift \citep[$z\sim0.3$,][]{Wold2017} contrasts with rest-frame EW measurements from other emission lines for galaxies at similar redshifts (e.g. H$\alpha$, {\sc [Oii]} and H$\beta$ + {\sc [Oiii]} EW$_0$) which are measured to be $\leq25$\,\AA\, at $z\sim0.3$, (e.g. SDSS: \citealt{Thomas2013}; HETDEX: \citealt{Adams2011}). It should be noted, however, that LAEs with very low EW$_0$ (down to $5$\,\AA) have been detected in some studies \citep[e.g.][]{Sobral2017,Arrabal2018}, highlighting the diversity of LAE populations. \cite{SobralMatthee2019} derived a simple empirical relation that estimates f$_{\rm esc, Ly\alpha}$ from EW$_0$: f$_{\rm esc, Ly\alpha}=0.0048\times$EW$_0$. This relation implies a connection between the intrinsic EW and the dust attenuation. A non-evolution of typical EW$_0$ with redshift could thus imply a non-evolution of f$_{\rm esc, Ly\alpha}$ in Ly$\alpha$-selected samples. A constant typical EW$_0=80$\,\AA\,across redshift would result in a typical f$_{\rm esc, Ly\alpha}\sim40\%$ for LAEs.

With the measurement of f$_{\rm esc, Ly\alpha}$ from EW$_0$, it is possible to derive the SFR of LAEs by translating Ly$\alpha$ flux into dust-corrected H$\alpha$ flux with simple assumptions. This provides a SFR computation which is independent of SED fitting and provides a comparison with SED-derived SFRs for LAEs even before observations with {\it James Webb Space Telescope}. Exploring how LAEs, which are typically low stellar mass galaxies, fit in the star formation ``Main Sequence" \citep[][]{Brinchmann2004,Noeske2007,Daddi2007,Schreiber2015} can shed light in a stellar mass range of the SFR-M$_\star$ relation which is still widely unconstrained at $z>2$. Previous studies have found that LAEs occupy the low stellar mass end of the Main Sequence at $z=2.5$ \citep[e.g.][]{Shimakawa2017} but also measured to be significantly above the Main Sequence extrapolation \citep{Whitaker2014} for low stellar masses at $z\sim2$ \citep[e.g.][]{Hagen2016,Kusakabe2018} and even at $z=4.9$ \citep{Harikane2018}. This suggests that LAEs are experiencing more intense star formation than the general population of galaxies of similar mass at similar redshifts, which may be explained by a burstier nature of star formation. We intend to expand these studies using a large sample of LAEs at $z\sim2-6$.

In this work, we use a uniformly selected sample of $\sim4000$ LAEs \citep[SC4K,][]{Sobral2018} to measure rest-frame UV properties  and their evolution from the end of reionisation at $z\sim6$ until the peak of star formation history at $z\sim2$. For our sample of galaxies, we measure EW$_0$, SFR, M$_\star$, UV luminosity (M$_{\rm UV}$) and UV continuum slope ($\beta$) for individual LAEs, using photometry measurements which we conduct ourselves, including data from UltraVISTA DR4, and by modelling SEDs using {\sc MAGPHYS} \citep{daCunha2008,daCunha2015}. Additionally, we discuss different approaches to measure SFR and how they influence our findings and we provide all our measurements in a public catalogue.

This paper is structured as follows: in Section \ref{sec:sample}, we present the SC4K sample of LAEs and detail how we conduct PSF aperture photometry and obtain SEDs and SED fits for each individual LAE. We present the properties of LAEs in Section \ref{sec:methods}, where we show the methodology we use to derive EW$_0$, SFR, M$_{\rm UV}$ and $\beta$. We present our results in Section \ref{sec:results}, looking into the M$_{\rm UV}$-$\beta$ and SFR-M$_\star$ relations and the potential evolution of EW$_0$ with redshift, along with physical interpretations. Finally, we present our conclusions in Section \ref{sec:conclusions}. Throughout this work, we use a $\Lambda$CDM cosmology with H$_0 = 70$\,km\,s$^{-1}$\,Mpc$^{-1}$, $\Omega_{\rm M} = 0.3$ and $\Omega _\Lambda = 0.7$. All magnitudes in this paper are presented in the AB system \citep{Oke1983} and we use a \citet{Chabrier2003} initial mass function (IMF) .

\section{Sample, photometry and SED fitting} \label{sec:sample}

\subsection{The sample: SC4K} \label{subsec:sample}

%
%
\begin{table*}
\setlength{\tabcolsep}{3pt}
\begin{center}
\caption{Overview of the SC4K sample of LAEs. We present the median of all measurements for each galaxy property, with the errors being the 16th and 84th percentile of the distribution. (1) LAE selection filter \citep{Sobral2018}; (2) Mean redshift of the sample based on Ly$\alpha$ within the filter FWHM; (3) Number of LAEs (Number of LAEs after removing sources with AGN signatures, see \S\ref{sec:sample_AGNs}); (4) Number of non-AGN LAEs with SEDs (percentage, see \S\ref{sec:number_seds}); (5) Ly$\alpha$ luminosity; (6) Ly$\alpha$ rest-frame EW; (7) SFR derived directly from L$_{\rm Ly\alpha}$ and EW$_{0}$ \citep[][see \S\ref{subsec:methods_SFR_Lya}]{SobralMatthee2019}; (8) Best likelihood SFR parameter from SED fitting; (9) Best likelihood stellar mass parameter from SED fitting; (10) UV magnitude computed by integrating the SED at $\lambda_0=1500$\,{\AA}, see \S\ref{subsec:muv}; (11) slope of the UV continuum measured from the SED fits, see \S\ref{subsec:beta}} \label{tab:overview}
\begin{tabular}{c | ccccccccccc}
\hline
(1)& (2) & (3) & (4) & (5) & (6) & (7) & (8) & (9) & (10) & (11)\\
\multicolumn{1}{c|}{Filter} &
\multicolumn{1}{c|}{Ly$\alpha\,z$} &
\multicolumn{1}{c|}{\# LAEs} &
\multicolumn{1}{c|}{\# SEDs} &
\multicolumn{1}{c|}{log$_{10}\,$L$_{\rm Ly\alpha}$} &
\multicolumn{1}{c|}{EW$_{0}$} &
\multicolumn{1}{c|}{SFR$_{\rm Ly\alpha}$} &
\multicolumn{1}{c|}{SFR$_{\rm SED}$} &
\multicolumn{1}{c|}{M$_{\star}$} &
\multicolumn{1}{c|}{M$_{\rm UV}$} &
\multicolumn{1}{c|}{$\beta$}\\
& & & (no AGN) & & (erg\,s$^{-1}$) & (\AA) & (M$_{\odot}$ yr$^{-1}$)& (log$_{10}\,$(M$_{\star}$/M$_{\odot}$)) & (AB) & \\
\hline
NB392 & 2.2 & 159 (137) & 129 (94\%) & $42.55^{+0.15}_{-0.15}$ & 79$^{+52}_{-44}$ & $4.7^{+4.9}_{-2.2}$ & $5.5^{+20.5}_{-3.6}$ & $9.5^{+0.5}_{-0.6}$ & $-19.6^{+1.0}_{-0.6}$ & $-1.8^{+0.9}_{-0.5}$ \\
IA427 & 2.5 & 741 (686) & 673 (98\%) & $42.64^{+0.22}_{-0.14}$ & 128$^{+220}_{-62}$ & $4.0^{+3.1}_{-1.8}$ & $2.9^{+6.9}_{-1.5}$ & $9.2^{+0.5}_{-0.5}$ & $-19.7^{+0.6}_{-0.6}$ & $-2.0^{+0.3}_{-0.4}$ \\
IA464 & 2.8 & 311 (284) & 283 (100\%) & $42.88^{+0.22}_{-0.15}$ & 121$^{+152}_{-52}$ & $6.8^{+4.6}_{-2.4}$ & $4.0^{+9.1}_{-1.6}$ & $9.1^{+0.6}_{-0.3}$ & $-20.2^{+0.5}_{-0.5}$ & $-2.1^{+0.5}_{-0.3}$ \\
IA484 & 3.0 & 711 (636) & 625 (98\%) & $42.83^{+0.18}_{-0.11}$ & 176$^{+340}_{-95}$ & $5.0^{+4.5}_{-2.0}$ & $3.1^{+5.8}_{-1.4}$ & $9.0^{+0.7}_{-0.3}$ & $-20.0^{+0.6}_{-0.7}$ & $-2.4^{+0.6}_{-0.0}$ \\
NB501 & 3.1 & 45 (38) & 31 (82\%) & $42.92^{+0.19}_{-0.13}$ & 170$^{+2259}_{-99}$ & $6.6^{+7.5}_{-3.2}$ & $6.2^{+15.2}_{-3.1}$ & $9.6^{+0.4}_{-0.5}$ & $-20.4^{+1.1}_{-0.8}$ & $-2.3^{+1.1}_{-0.2}$ \\
IA505 & 3.2 & 483 (437) & 433 (99\%) & $42.89^{+0.19}_{-0.13}$ & 142$^{+351}_{-71}$ & $6.3^{+4.9}_{-2.5}$ & $4.5^{+6.5}_{-2.0}$ & $9.4^{+0.5}_{-0.5}$ & $-20.2^{+0.6}_{-0.6}$ & $-2.1^{+0.4}_{-0.4}$ \\
IA527 & 3.3 & 641 (593) & 573 (97\%) & $42.84^{+0.19}_{-0.10}$ & 149$^{+245}_{-74}$ & $5.7^{+5.1}_{-2.3}$ & $4.1^{+5.7}_{-1.9}$ & $9.4^{+0.6}_{-0.6}$ & $-20.2^{+0.5}_{-0.6}$ & $-2.0^{+0.3}_{-0.5}$ \\
IA574 & 3.7 & 98 (88) & 87 (99\%) & $42.98^{+0.14}_{-0.13}$ & 97$^{+72}_{-39}$ & $10.9^{+6.4}_{-4.9}$ & $6.7^{+6.9}_{-2.7}$ & $9.3^{+0.7}_{-0.2}$ & $-20.8^{+0.5}_{-0.4}$ & $-2.4^{+0.8}_{-0.0}$ \\
IA624 & 4.1 & 142 (139) & 116 (83\%) & $43.02^{+0.18}_{-0.06}$ & 186$^{+666}_{-99}$ & $6.7^{+8.2}_{-1.8}$ & $6.1^{+9.1}_{-2.8}$ & $9.2^{+0.5}_{-0.5}$ & $-20.5^{+0.5}_{-0.6}$ & $-1.9^{+0.3}_{-0.5}$ \\
IA679 & 4.6 & 79 (75) & 69 (92\%) & $43.25^{+0.15}_{-0.05}$ & 186$^{+267}_{-89}$ & $11.6^{+12.2}_{-2.8}$ & $9.3^{+18.6}_{-4.0}$ & $9.5^{+0.8}_{-0.3}$ & $-21.2^{+0.6}_{-0.5}$ & $-2.4^{+0.8}_{-0.0}$ \\
IA709 & 4.8 & 81 (77) & 73 (95\%) & $43.16^{+0.13}_{-0.10}$ & 124$^{+200}_{-56}$ & $13.2^{+9.9}_{-5.5}$ & $9.1^{+15.8}_{-3.8}$ & $9.4^{+0.5}_{-0.3}$ & $-21.1^{+0.5}_{-0.4}$ & $-2.0^{+0.3}_{-0.5}$ \\
NB711 & 4.8 & 78 (74) & 56 (76\%) & $42.74^{+0.28}_{-0.16}$ & 80$^{+64}_{-42}$ & $7.8^{+11.2}_{-3.6}$ & $14.4^{+61.0}_{-9.5}$ & $9.7^{+0.6}_{-0.6}$ & $-20.9^{+0.5}_{-0.8}$ & $-1.9^{+0.8}_{-0.5}$ \\
IA738 & 5.1 & 79 (75) & 65 (87\%) & $43.25^{+0.17}_{-0.14}$ & 120$^{+222}_{-47}$ & $15.7^{+15.5}_{-7.6}$ & $16.0^{+32.4}_{-9.2}$ & $9.6^{+0.7}_{-0.3}$ & $-21.3^{+0.4}_{-0.7}$ & $-1.8^{+0.2}_{-0.6}$ \\
IA767 & 5.3 & 33 (30) & 29 (97\%) & $43.37^{+0.20}_{-0.07}$ & 134$^{+169}_{-48}$ & $18.7^{+15.0}_{-7.4}$ & $20.6^{+50.5}_{-10.8}$ & $9.7^{+0.3}_{-0.4}$ & $-21.6^{+0.4}_{-0.5}$ & $-2.0^{+0.3}_{-0.4}$ \\
NB816 & 5.7 & 192 (186) & 108 (58\%) & $42.82^{+0.27}_{-0.11}$ & 235$^{+547}_{-169}$ & $5.2^{+6.4}_{-2.4}$ & $28.5^{+83.7}_{-20.8}$ & $9.9^{+0.4}_{-0.5}$ & $-21.4^{+0.6}_{-0.6}$ & $-1.8^{+0.7}_{-0.6}$ \\
IA827 & 5.8 & 35 (35) & 27 (77\%) & $43.44^{+0.19}_{-0.11}$ & 325$^{+963}_{-266}$ & $22.0^{+47.5}_{-8.4}$ & $25.3^{+80.1}_{-16.1}$ & $9.9^{+0.6}_{-0.4}$ & $-22.0^{+0.8}_{-1.0}$ & $-1.8^{+0.7}_{-0.6}$ \\
\hline
GLOBAL & 4.1 & 3908 (3590) & 3377 (94\%) & $42.84^{+0.27}_{-0.17}$ & 138$^{+281}_{-70}$ & $5.9^{+6.3}_{-2.6}$ & $4.4^{+10.5}_{-2.4}$ & $9.3^{+0.6}_{-0.5}$ & $-20.2^{+0.7}_{-0.8}$ & $-2.1^{+0.5}_{-0.4}$ \\
\hline
\end{tabular}
\end{center}
\end{table*}

%
%
\begin{figure*}
  \centering
  \includegraphics[width=\textwidth]{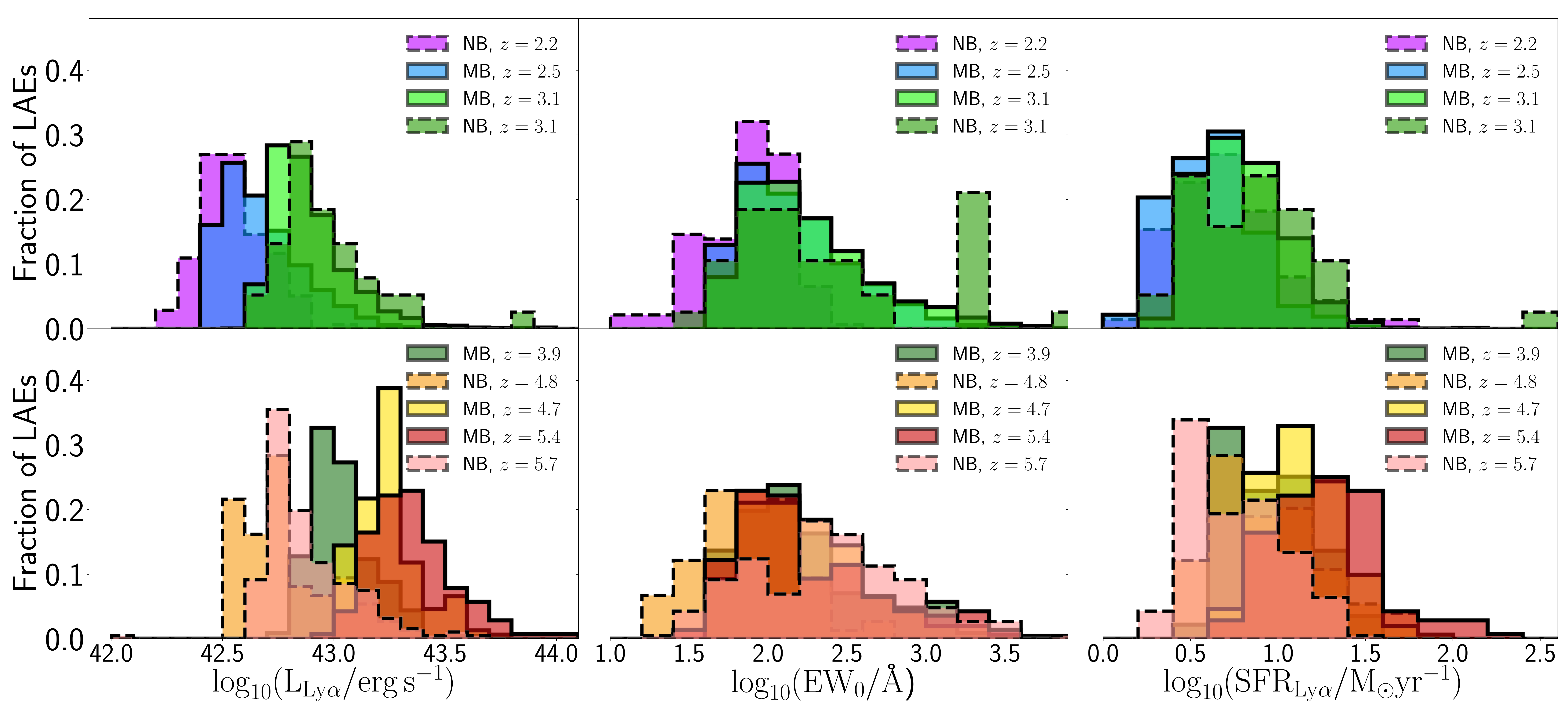}
  \caption{Distributions of parameters derived directly from photometry. Ly$\alpha$ luminosity (left panel), EW$_0$ (middle panel) and SFR derived directly from L$_{\rm Ly\alpha}$ and EW$_{0}$ \citep[][see \S\ref{subsec:methods_SFR_Lya}; right panel]{SobralMatthee2019}. MB (NB) data are shown as filled (dashed) lines. For each parameter, top panels show the $z\leq3.1$ sample and bottom panels show the higher redshift LAEs. The EW$_0$ peak at $z=3.1$ (NB) is artificial and it is the upper limit of the EW$_0$, obtained from the flux upper limit. AGN have been removed.}
  \label{fig:histograms1}
\end{figure*}

We use the public SC4K sample of LAEs \citep[Slicing COSMOS with 4k LAEs,][]{Sobral2018}, which contains 3908 sources selected due to their high Ly$\alpha$ EW at $z\sim2-6$. These LAEs were selected with wide field surveys conducted with Subaru and the Isaac Newton telescopes, using 16 (12+4) medium+narrow bands (MB+NB) over 2 deg$^2$ in the COSMOS field \citep{Capak2007,Scoville2007,Taniguchi2015}, covering a full comoving volume of $\sim10^8$\,Mpc$^3$. For full details on the selection of the sample see \cite{Sobral2018}. Briefly, the selection criteria applied were i) EW$_0$ cut of 50\,\AA\ for MBs, 25\,\AA\, for NBs and 5\,\AA\, for the NB at $z=2.23$: see \citealt{Sobral2017}); ii) significant excess emission in the selection medium/narrow band, $\Sigma>3$ \citep[see][]{Bunker1995,Sobral2013}; iii) colour break blueward of the detected Ly$\alpha$ emission, due to the expected presence of a Lyman Break; iv) removal of sources with strong red colours which are typically lower redshift contaminants where the Balmer break mimics a Lyman break; v) visual inspection of all candidates to remove spurious sources and star artefacts.

We show an overview of the properties of the SC4K LAEs, split by selection bands, in Table \ref{tab:overview}. For each selection band, we provide the median of each property and the 16th (84th) percentiles of its distribution as lower (upper) uncertainties. Additionally, in Fig. \ref{fig:histograms1} we show a histogram distribution of Ly$\alpha$ luminosity (L$_{\rm Ly\alpha}$), EW$_0$ (see \S\ref{subsec:EW_Lya}) and SFR using the \cite{SobralMatthee2019} calibration (see \S\ref{subsec:methods_SFR}). The differences in the lower end distribution of L$_{\rm Ly\alpha}$ are driven by an increasing luminosity distance and a roughly similar flux limit. The evolution of the Ly$\alpha$ luminosity function is presented in \cite{Sobral2018}. 

We note that extensive analysis of the SC4K public sample have already been conducted in previous works. For example, \cite{PaulinoAfonso2018} studied the UV morphologies of the sample and found that UV sizes of LAEs are constant from $z\sim2$ to $z\sim6$ with effective radii sizes of r$_e\sim1.0 \pm 0.1$ kpc. \cite{Shibuya2019} analysed the radial surface brightness profiles of $\sim9000$ LAEs (including SC4K) and found that LAEs typically have small sizes, similar to those presented by \cite{PaulinoAfonso2018}. This means SC4K LAEs are unresolved in the continuum in ground-based data. \cite{Khostovan2019} derived clustering properties of the sample and measured typical halo masses of $\sim10^{11}$\,M$_\odot$ in NB-selected LAEs and $\sim10^{11}-10^{12}$\,M$_\odot$ in MB-selected LAEs, showing the clustering and typical dark matter halo masses that host LAEs is strongly dependent on L$_{\rm Ly\alpha}$. They find more luminous LAEs reside in more massive dark matter haloes. \cite{Calhau2019} study the X-ray and radio properties of the SC4K sample, estimating black hole accretion rates which can reach $\sim3$\,M$_\odot$ yr$^{-1}$ in the most extreme sources. They also find that the overall AGN fraction of LAEs is low ($<10\%$) but dependent on L$_{\rm Ly\alpha}$, significantly increasing with increasing luminosity and approaching $100\%$ at L$_{\rm Ly\alpha}>10^{44}$ erg s$^{-1}$.

\subsubsection{X-ray and radio AGN in SC4K} \label{sec:sample_AGNs}

In total we have 3908 LAEs in our sample, with 254 detected in X-ray and 120 detected in radio (56 in both), resulting in 318 AGN candidates \citep{Calhau2019}. LAEs which are detected in the X-ray and/or radio are classified as AGN as star-forming processes would require ${\rm SFR}\gtrsim1000$\,M$_\odot$ yr$^{-1}$ to be detected above the flux limit at such wavelengths and redshifts \citep[see discussion in][]{Calhau2019}. The number of AGNs reported in this paper constitutes an extra 177 sources compared to the ones originally reported in \cite{Sobral2018}, with the additional sources being identified by reaching lower S/N with deep {\it Chandra} data \citep[COSMOS Chandra Legacy, ][]{Civano2016} and VLA radio data at 1.4 GHz \citep[VLA-COSMOS Survey,][]{Schinnerer2004,Schinnerer2007,Bondi2008,Schinnerer2010} and by including 3 GHz radio data \citep{Smolcic2017}. We note, however, that due to available coverage, \cite{Calhau2019} only probe 3705 SC4K LAEs with X-Ray and radio data. Throughout this work, SC4K AGNs may be shown in figures (clearly highlighted as such) but are removed from any fitting/binning and median values in tables unless stated otherwise as we focus on the properties of the star-forming population. The catalogue that is provided in this paper has a flag for sources detected in X-Ray and radio (see \S\ref{sec:catalogue}).

\subsubsection{Redshift binning}

To improve the S/N in certain redshift ranges and for clearer visualisation of results, we frequently group multiple MB filters in specific redshift bins throughout this paper, following the same grouping scheme as \cite{Sobral2018}: $z=2.5\pm0.1$ (IA427), $z=3.1\pm0.4$ (IA464, IA484, IA505, IA527); $z=3.9\pm0.3$ (IA574, IA624); $z=4.7\pm0.2$ (IA679, IA709); $z=5.4\pm0.5$ (IA738, IA767, IA827). We generally study the NBs separately as there are some relevant distinctions between MBs and NBs, most significantly the flux limit and EW$_0$ cut. Additionally, analysing the two separately provides independent results and allows checks for systematics.

\subsection{Multi-wavelength data}

We use the extensive archive of publicly available multi-wavelength data in the COSMOS field to conduct accurate photometric measurements in the UV, optical, near-infrared (NIR), mid-infrared (MIR) and FIR wavelengths for each SC4K LAE, individually. A summary of the filters used, effective wavelength, width and limiting magnitude is provided in Table \ref{tab:filters}. We use optical broad band (\textit{B, V, g$^+$, r$^+$, i$^+$, z$^{++}$}), medium band (IA427, IA464, IA484, IA505, IA527, IA574, IA624, IA679, IA709, IA738, IA767, IA827) and narrow band (NB711, NB816) data taken with the Subaru/SuprimeCam \citep{Taniguchi2007,Capak2007}, retrieved from the COSMOS Archive\footnote{https://irsa.ipac.caltech.edu/data/COSMOS/images/}. Additionally, we use the $u$ band from CFHT/MegaCam. We use deep NIR data (\textit{Y, J, H, Ks}) from UltraVISTA DR4 \citep{McCracken2012}, taken with VISTA/VIRCAM \citep{Sutherland2015}. Data used have a 0.15$"$\,pix$^{-1}$ pixel scale and are calibrated to a zero-point of 31.4 mag (30 mag for UltraVISTA and $u$ images). For MIR coverage, we use data from {\it Spitzer}/IRAC, channels 1 (3.6$\micron$) and 2 (4.5$\micron$) from SPLASH \citep{Steinhardt2014} and channels 3 (5.6$\micron$) and 4 (8.0$\micron$) from S-COSMOS \citep{Sanders2007}. IRAC data have a zero-point of 21.5814 mag and a pixel scale of 0.6$"$\,pix$^{-1}$.

For the FIR coverage, we use 100$\micron$ and 160$\micron$ data \citep[PEP, ][]{Lutz2011} taken with {\it Herschel}/PACS \citep{Pilbratt2010} and 250$\micron$, 350$\micron$ and 500$\micron$ data \citep[HerMES, ][]{Griffin2010, Oliver2012} taken with {\it Herschel}/SPIRE. The five listed FIR images have a pixel scale of 1.2$"$\,pix$^{-1}$, 2.4$"$\,pix$^{-1}$, 6$"$\,pix$^{-1}$, 8.3$"$\,pix$^{-1}$ and 12$"$\,pix$^{-1}$, respectively. FIR images are calibrated to provide fluxes in Jansky and thus have a zero-point of 8.9 mag.

%
%
\begin{table*}
\caption{Overview of the photometric filters used in this work ranked from the lowest to highest wavelengths. (1) Photometric filter; (2) Effective wavelength; (3) Filter FWHM; (4) 3\,$\sigma$ magnitude depth measured in a fixed 2$"$ aperture; (5) Correction term summed to the measured magnitudes to correct for systematic offsets ($^\star$includes an additional offset to correct the systematic uncertainties $\S$\ref{subsec:sf}; $^\dagger$denotes values obtained from the deblended FIR catalogue presented by \citealt{Jin2018}); (6) Filter dependent dust correction that is subtracted from the measured magnitudes; (7) Instrument and telescope used for the observations; (8) Source of the data.} \label{tab:filters}
\begin{center}
\begin{tabular}{c | ccccccccc}
\hline
\multicolumn{1}{c|}{Filter} &
\multicolumn{1}{c|}{$\lambda_{\rm eff}$} &
\multicolumn{1}{c|}{FWHM} &
\multicolumn{1}{c|}{Depth} &
\multicolumn{1}{c|}{$s_f$} &
\multicolumn{1}{c|}{$A_\lambda$} &
\multicolumn{1}{c|}{Instrument, Telescope} &
\multicolumn{1}{c|}{Source}\\
& (\AA) & (\AA) & ($3\,\sigma$, 2$"$) &&&\\
(1) & (2) & (3) & (4) & (5) & (6) & (7) & (8) \\
\hline
$u$  & 3911.0 & 538.0 & 27.8 & 0.054 & 0.0878 & MegaCam, CFHT & \citet{Capak2007}\\
$IA427$  & 4256.3 & 206.5 & 27.0 & 0.037 & 0.0816 & Suprime-Cam, Subaru & \citet{Capak2007}\\
$B$  &     4439.6 & 806.7 & 28.3 & -0.242 & 0.0784 & Suprime-Cam, Subaru & \citet{Capak2007}\\
$IA464$   & 4633.3 & 218.0 & 26.9 & 0.013  & 0.0750 & Suprime-Cam, Subaru & \citet{Capak2007}\\
$g^+$   & 4728.3 & 1162.9 & 27.6 & 0.024 & 0.0733 & Suprime-Cam, Subaru & \citet{Capak2007}\\
$IA484$  & 4845.9 & 228.5 & 27.0 & 0.000 & 0.0713 & Suprime-Cam, Subaru & \citet{Capak2007}\\
$IA505$   &5060.7 & 230.5 & 26.8 & -0.002 & 0.0678 &  Suprime-Cam, Subaru & \citet{Capak2007}\\
$IA527$   &5258.9 & 242.0 & 27.1 & 0.026 & 0.0646 & Suprime-Cam, Subaru & \citet{Capak2007}\\
$V$   &    5448.9 & 934.8 & 27.6 & 0.046$^\star$ & 0.0616 &  Suprime-Cam, Subaru & \citet{Capak2007}\\
$IA574$  & 5762.1 & 271.5 & 26.8 & 0.078 & 0.0570 & Suprime-Cam, Subaru & \citet{Capak2007}\\
$IA624$ & 6230.0 & 300.5 & 26.8 & 0.002 & 0.0506 & Suprime-Cam, Subaru & \citet{Capak2007}\\
$r^+$    &   6231.8 & 1348.8 & 27.7 & 0.003 & 0.0506 &  Suprime-Cam, Subaru & \citet{Capak2007}\\
IA679  & 6778.8 & 336.0 & 26.7 & 0.039$^\star$ & 0.0442 & Suprime-Cam, Subaru & \citet{Capak2007}\\
IA709  & 7070.7 & 315.5 & 26.8  & -0.024 & 0.0411 &  Suprime-Cam, Subaru & \citet{Capak2007}\\
NB711  & 7119.6 & 72.5 & 25.9 & 0.014 & 0.0406 & Suprime-Cam, Subaru & \citet{Capak2007}\\
IA738  & 7358.7 & 323.5 & 26.5 & 0.017 & 0.0383 & Suprime-Cam, Subaru & \citet{Capak2007}\\
i$^+$    &   7629.1 & 1489.4 & 27.2 & 0.019 & 0.0360 & Suprime-Cam, Subaru & \citet{Capak2007}\\
IA767  & 7681.2 & 364.0 &  26.5 & 0.041 & 0.0356 & Suprime-Cam, Subaru & \citet{Capak2007}\\
NB816  & 8149.0 & 119.5 &  26.6  & 0.068 & 0.0320 & Suprime-Cam, Subaru & \citet{Capak2007}\\
IA827  & 8240.9 & 343.5 &  26.5 & -0.019  & 0.0313 & Suprime-Cam, Subaru & \citet{Capak2007}\\
z$^{++}$  &  9086.6 & 955.3 & 26.8 & -0.037 & 0.0265 & Suprime-Cam, Subaru & \citet{Capak2007}\\
  Y    &    10211.2  & 930.0 & 26.2  & 0.0 & 0.0211 & VIRCAM, VISTA & \citet{McCracken2012} (DR4)\\
  J    &     12540.9 & 172.0 & 25.8  & 0.0 & 0.0144 & VIRCAM, VISTA & \citet{McCracken2012} (DR4)\\ 
  H    &    16463.7 & 2910 &  26.1 & 0.0 & 0.0088 & VIRCAM, VISTA & \citet{McCracken2012} (DR4)\\
  Ks    &     21487.7 & 3090 &  25.8 & 0.0& 0.0053 & VIRCAM, VISTA & \citet{McCracken2012} (DR4)\\ 
  $IRAC1$ &  35262.5 & 7412 &  25.6 & 0.002 & 0.0021 & IRAC, {\it Spitzer} & \citet{Steinhardt2014}\\
  $IRAC2$ &   44606.7 & 10113 & 25.5 & 0.000 & 0.0014 & IRAC, {\it Spitzer} & \citet{Steinhardt2014} \\
  $IRAC3$ &   56764.4 & 13499 & 22.6 & 0.013 & 0.0010 & IRAC, {\it Spitzer} & \citet{Sanders2007}\\
  $IRAC4$ &   77030.1 & 28397 & 22.5  & -0.171 &  0.0007 & IRAC, {\it Spitzer} & \citet{Sanders2007}\\
  100$\micron$    &   979036.1 & 356866 & 15.4 & 0.20$^\dagger$& 0.0000 & PACS, {\it Herschel} & \citet{Lutz2011}\\
  160$\micron$    &   1539451.3  & 749540 & 14.3 &-0.06$^\dagger$ & 0.0000  & PACS, {\it Herschel} & \citet{Lutz2011}\\
  250$\micron$    &   2471245.1  & 658930 & 10.9 &-0.49$^\dagger$ &  0.0000 & SPIRE, {\it Herschel} &  \citet{Oliver2012}\\
  350$\micron$    &   3467180.4 & 937200 & 10.6 &-0.15$^\dagger$ &  0.0000 & SPIRE, {\it Herschel} &  \citet{Oliver2012}\\
  500$\micron$    &   4961067.7 & 1848042 & 10.6 &0.03$^\dagger$ &  0.0000 & SPIRE, {\it Herschel} &  \citet{Oliver2012}\\
\hline
\end{tabular}
\end{center}
\end{table*}

\subsection{Multi-wavelength photometry} \label{sec:photometry}

Accurate photometric measurements are essential to obtain robust SEDs and derive accurate galaxy properties, particularly for sources that are faint in the continuum. While there is a plethora of publicly available catalogues for the COSMOS field \citep[e.g.][]{Ilbert2009,Laigle2015}, such catalogues are typically broad band selected and thus miss a significant number of line-emitters, especially faint, high EW sources. For example, 9\% of our LAEs are not detected in the $i$ band-selected catalogue from \cite{Ilbert2009} with 1$"$ radius matching and 29\% of SC4K LAEs are not detected in the NIR-selected catalogue from \cite{Laigle2015}. Continuum faint sources with very blue UV continuum slopes have low fluxes in the observed optical and will fall below the detection thresholds of NIR selected catalogues \citep[e.g.][]{Laigle2015}, particularly if they have low stellar masses. Therefore, to obtain consistent, controllable and uniform measurements for the entire sample of LAEs, we conduct our own aperture photometry and estimate errors locally using empty apertures. We also compare our photometry with measurements from the COSMOS catalogues and find a very good agreement. Furthermore, because we have measured the sizes in the rest-frame UV and found SC4K LAEs to be very compact (point-like for the data we use; r$_e=1.0$ kpc corresponds to 0.13$"$ at $z=3$), we opt to conduct PSF photometry, as fully explained in \S\ref{subsec:aper_photo}.

\subsection{Aperture photometry of SC4K LAEs} \label{subsec:aper_photo}

\subsubsection{Overview of our aperture photometry} \label{subsec:overview_aper}

In order to obtain accurate PSF aperture photometry for individual LAEs, for each band, we estimate the total magnitude by following the steps:
\begin{itemize}
\item conducting photometry in fixed apertures (\S\ref{subsec:fixed_aper});
\item applying aperture corrections based on PSF stars around each LAE (\S\ref{subsec:psf});
\item applying reddening corrections (\S\ref{subsec:extinction});
\item introducing systematic offset corrections based on known offsets and COSMOS catalogues (\S\ref{subsec:sf});
\end{itemize}

Magnitudes per source and per band are computed as:
\begin{equation} \label{eq:mag}
{\rm mag=mag_0+aper_{cor}}+s_f-A_\lambda,
\end{equation}
where mag$_0$ is the magnitude calculated by converting the flux obtained in fixed apertures (typically 2$"$ diameter for most of the data) to the AB magnitude system before any correction is applied, aper$_{\rm cor}$ is the aperture correction derived per band and per source, based on PSF stars around each LAE, $s_f$ the systematic offset correction for the filter and $A_\lambda$ the reddening correction computed for the effective wavelength of the filter. The error in the final magnitude is obtained by propagating the error in flux, scaling the error with the correction that was applied to the flux and then adding 30\% of the total correction to the error in flux\footnote{We note that we use 30\% as a conservative approach to add unknown systematic errors.}. Aperture photometry in the FIR is discussed separately in $\S$\ref{subsec:fir_photo}.

\subsubsection{Aperture photometry in fixed apertures} \label{subsec:fixed_aper}

We conduct aperture photometry centred on the position of each SC4K LAE \citep{Sobral2018} over all the filters listed in Table \ref{tab:filters}. We do this by creating 200x200 pixel (30$"$$\times$30$"$ for a 0.15$"$\,pix$^{-1}$ pixel scale) cutouts, where we conduct the photometry\footnote{We use PSF stars beyond this region.}. For optical to MIR images, we use 2$"$ diameter apertures. We estimate the background noise by placing 2000 2$"$ apertures in random positions of the field where there are no detections above 2\,$\sigma$ (given by the segmentation maps per filter produced by SExtractor; \citealt{Bertin1996}) and subtract it from the counts of the aperture placed on the LAE. Upper and lower errors are measured as the 84th and 16th percentiles of all random apertures. We repeat this procedure per band per source.

\subsubsection{Aperture correction} \label{subsec:psf}

The original point-spread function (PSF) was kept across all images as we have opted for correcting the photometry with PSF stars, instead of PSF matching the data, in order to avoid modifying the data and confuse nearby sources. Fixed aperture photometry in non-PSF matched images requires correction of the PSF effect on photometry so we can obtain total fluxes and total magnitudes for point-like sources. To do this, we measure the magnitude of stars\footnote{Selected from \cite{Ilbert2009}: photoz=0.0; stellaricity=1; detected in the point-source catalogue 2MASS \cite{Skrutskie2006}; visually checked to remove binary systems or close projections.} in 2$"$ apertures and with {\sc mag\_auto} \citep{Bertin1996}\footnote{aper$_{\rm cor}$ = {\sc mag\_auto}$-$mag$_0$.}. We define the correction factor (aper$_{\rm cor}$ in Equation \ref{eq:mag}) as the difference between {\sc mag\_auto} and magnitudes measured in 2$"$ apertures. This correction is valid for point-like sources, an assumption that should be valid for our LAEs given the rest-frame UV sizes as measured by \cite{PaulinoAfonso2018} using high-resolution {\it HST}/ACS images. The correction term is measured for each filter, and it is the median correction of stars within a 0.3 degree radius around each LAE, accounting for spatial variations of the PSF per band.

\subsubsection{Galactic extinction correction} \label{subsec:extinction}

We correct for dust attenuation along the line-of-sight due to our Galaxy. For the COSMOS field, the median galactic extinction is measured to be $E(B-V)=0.0195\pm0.006$ \citep{Capak2007}. The slope of the extinction curve with wavelength is parametrised by the factor $R(V)$:
\begin{equation}
R(V)\equiv\frac{A(V)}{E(B-V)},
\end{equation}
where $A(V)$ is the total extinction at the V band. For the diffuse interstellar medium, the median value of $R(V)$ is estimated to be 3.1 \citep[e.g.][]{Fitzpatrick1999} and it is the value used in this paper. We use the model from \cite{Fitzpatrick2007} where the attenuation at a wavelength ($\lambda$) becomes:
\begin{equation}
A_\lambda=A(V)\left(1+\frac{k}{R(V)}\right),
\end{equation}
where $k$ is a polynomial expansion of $\lambda^{-1}$ (Equation 2 from \citealt{Fitzpatrick2007}) with a linear component for UV wavelengths, a curvature term for the far-UV and a Lorentzian-like bump at 2175\,\AA. We determine $A_\lambda$ for the effective wavelength of each filter and show its value for each filter in Table \ref{tab:filters}.

%
%
\begin{figure*}
\begin{tabular}{ll}
  \centering
  \includegraphics[width=0.487\textwidth]{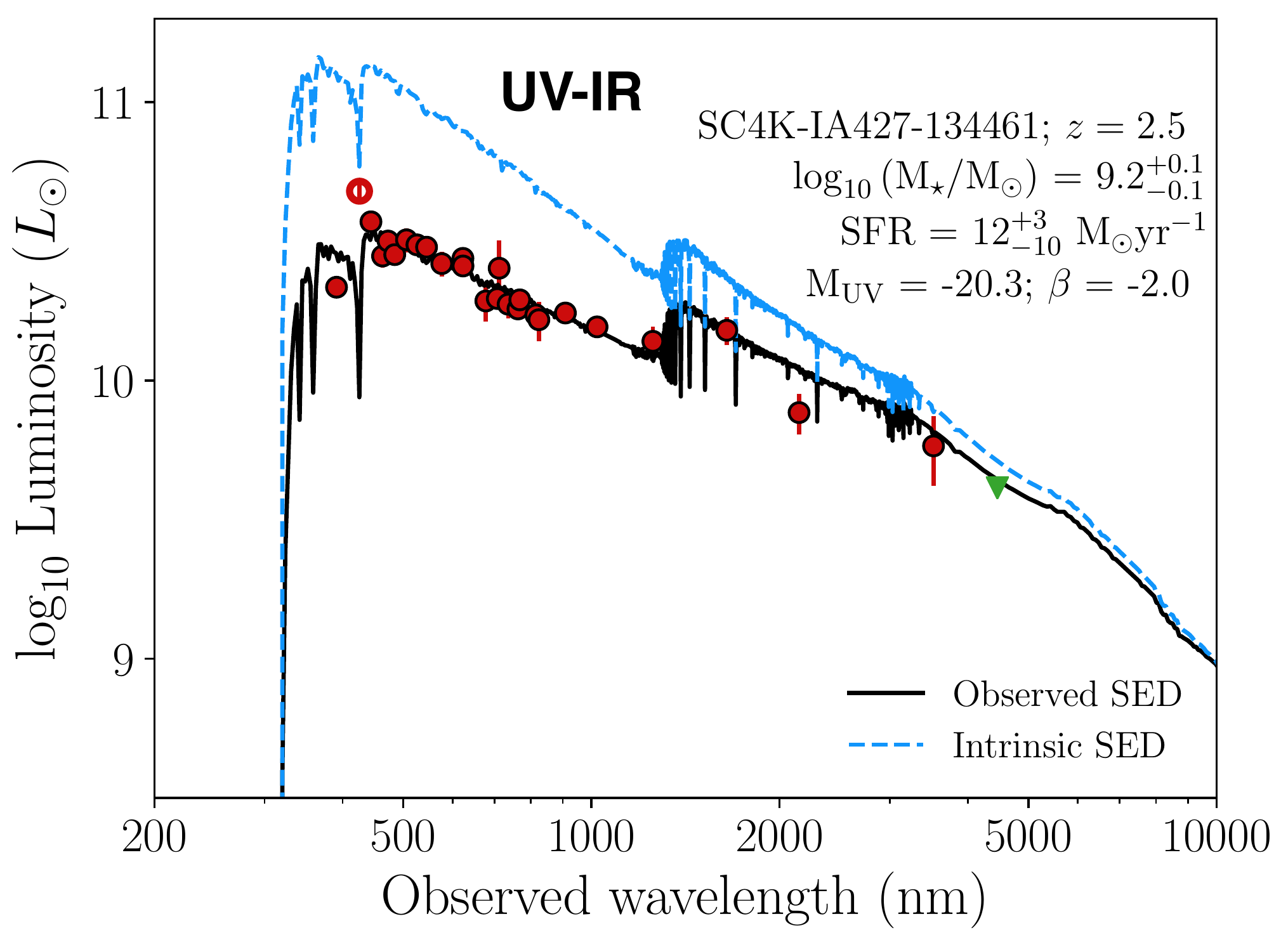}
  &
  \includegraphics[width=0.487\textwidth]{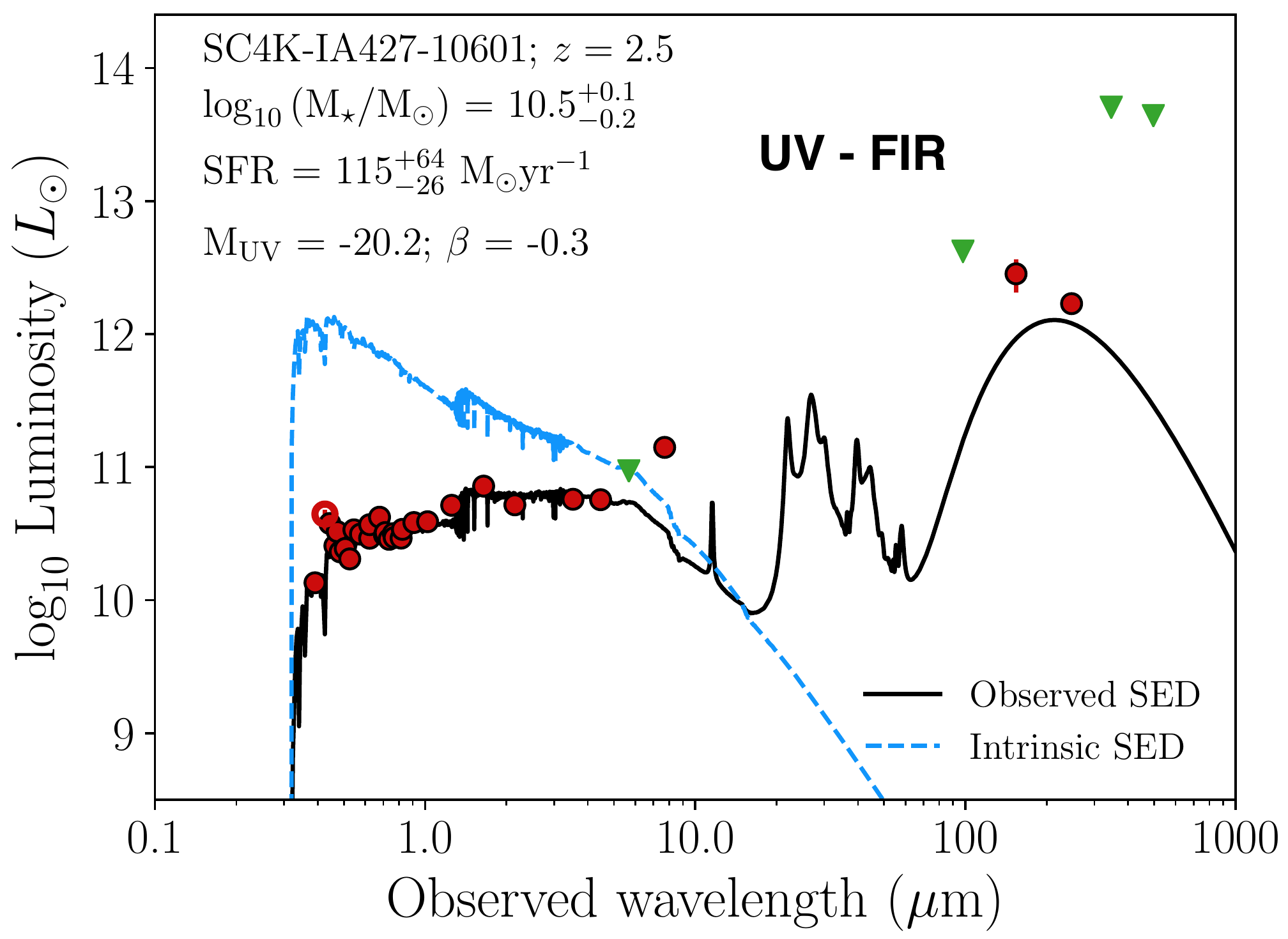}  
\end{tabular}
  \caption{{\it Left:} SED of SC4K-IA427-134461 (at $z=2.5$), for observed UV-IR wavelengths as we only obtain upper limits in the FIR. Red circles show the luminosity (in solar units) measured at the corresponding observed wavelength and green arrows show the upper limits for non-detections, where the flux is $<3\,\sigma$. Unfilled circles are the luminosity at the NB/MB where the LAE was selected, and we note that this filter was not used to derive the SED fit. The black line is the best-fit SED to the observed photometry and the blue dashed line the intrinsic (dust-free) SED. This is an example of a very blue ($\beta=-2.0$) and low stellar mass (M$_{\star}=10^{9.2}$\,M$_\odot$) LAE. {\it Right:} Same as left panel but for SC4K-IA427-10601 (at $z=2.5$) and at a wider wavelength range, showing FIR wavelengths as this LAE is detected in 250$\micron$ and 350$\micron$ due to the presence of dust. This LAE is redder ($\beta=-0.3$) and more massive (M$_{\star}=10^{10.5}$\,M$_\odot$). Note that this LAE is not representative of the SC4K sample as only $\sim3\%$ (2\%) non-AGN LAEs are as massive (as red).}
  \label{fig:SED}
\end{figure*}
 
\subsubsection{FIR photometry} \label{subsec:fir_photo}

For FIR data, due to the large PSF of 7.2$"$, 12$"$, 18.15$"$, 25.15$"$ and 36.30$"$ (100$\micron$, 160$\micron$, 250$\micron$, 350$\micron$ and 500$\micron$, respectively), the usage of 2$"$ diameter aperture photometry is not viable. We conduct PSF aperture photometry using apertures which are the size of the PSF: radius of 6, 5, 3, 3 and 3 pixels, respectively (retrieving 67\% of the total flux), with the same random empty aperture procedure to estimate background. This allows us to then apply aperture corrections of 1/0.67 to get full fluxes for point-like sources. For 100$\micron$ (160$\micron$), we multiply the flux by the filter correction factor 1.1 (1.2) as described in the PEP public data release notes \citep[see][]{Lutz2011}.

However, the blending of sources is still a serious issue, as the large pixel scale makes it difficult to establish if a detection is produced by one of our LAEs or by a neighbouring source. To solve this, we use the FIR measurements from the publicly available deblended COSMOS catalogue \citep{Jin2018}, where FIR emission is deblended to match optical-NIR coordinates. With a 1$"$ match to the deblended catalogue, there are 14, 11, 29, 19 and 12 SC4K LAEs with 3\,$\sigma$ detections in 100$\micron$, 160$\micron$, 250$\micron$, 350$\micron$ and 500$\micron$, respectively. Whenever a source is undetected in the FIR, we assign the local estimate of the background as an upper limit, which we measure with 2000 empty apertures the size of the PSF. We ensure our own flux measurements are consistent with \cite{Jin2018} (see $\S$\ref{subsec:sf}).

\subsubsection{Systematic offsets} \label{subsec:sf}

We correct for systematic offsets ($s_f$) in the photometry by applying the corrections derived by \cite{Ilbert2009} (we present these values in Table \ref{tab:filters}). After applying the systematic offsets and all previous correction terms, we compare our total magnitudes with measurements from \cite{Ilbert2009} and \cite{Laigle2015}. We find no statistically significant difference with our measurements except for two filters (IA679, V) which have systematic offsets of $\sim0.5$ mag. We apply a further correction (included in the $s_f$, Table \ref{tab:filters}) to our magnitudes, so the median of the magnitude difference becomes zero. For FIR magnitudes, we estimate the systematic correction term from the FIR deblended catalogue \citep{Jin2018}, also presented in Table \ref{tab:filters}.

\subsection{Spectral Energy Distributions of SC4K LAEs} \label{subsec:methods_SED}

Having conducted photometry in the 34 filters listed in Table \ref{tab:filters}, we can now explore the SED of each individual LAE, observed from UV to FIR. We use the publicly available SED-fitting code {\sc MAGPHYS}\footnote{http://www.iap.fr/magphys/} \citep{daCunha2008,daCunha2012} with the high-redshift extension \cite[see][]{daCunha2015}, to obtain SED fits for each individual galaxy, using our rest-frame UV, optical and NIR-FIR photometric measurements.

{\sc MAGPHYS} is based on dust attenuation models from \citet{Charlot2000} and uses the stellar population synthesis model from \citet{Bruzual2003} with a \citet{Chabrier2003} IMF (range $0.1-100$ M$_\odot$) to compute the emission of simple stellar populations (SSPs, populations of coeval stars with similar properties). We use the prescription of \cite{Madau1995} to model the Intergalactic Medium (IGM). The software generates a library of model SEDs for galaxies at the mean redshift of the NB/MB filter (see Table \ref{tab:overview}) and for the given photometric bands. The modelled SED of a galaxy is composed by the weighted sum of SSPs, with the star formation history (SFH) being a continuously delayed exponential function with an early rise followed by a decay. Instantaneous bursts of star formation of random duration (lasting 30-300 Myr) and amplitude (forming mass between 0.1-100 times the mass formed by the continuous SFH) are superimposed. A Bayesian approach is then used to compare model SEDs with observed photometry, creating a parameter likelihood distribution for several galaxy properties such as stellar mass, SFR and dust attenuation.

As the models are purely stellar (no nebular line fitting), we do not fit photometry from filters where we expect strong nebular emission, namely Ly$\alpha$ at the selection NB or MB filters, as it is by definition significant in our Ly$\alpha$-selected sample. While we do not remove photometry from filters which may have contribution from other emission lines such as H$\alpha$ (IRAC filters at $z\sim4-6$) or {\sc [Oiii]} (H-K bands at $z\sim2-3$), by removing the Ly$\alpha$-contaminated filter, combined with the large number of filters used, we do not expect an overestimation of masses due to nebular line contamination. We explore this by rerunning {\sc MAGPHYS} for the entire $z=2.5$ sample (IA427) after removing the $H$ and $K$ bands, which may be contaminated from {\sc [Oiii]} and H$\alpha$ emission, respectively, and compare the difference of estimated stellar masses. We find that when removing both $H$ and $K$, the median difference of stellar masses is 0, with no dependence on mass, and the average difference -0.07 (M$_{\rm \star, no\,HK}$ being slightly smaller). Removing $H$ and $K$ makes the estimation of stellar mass more uncertain as the rest-frame optical becomes more poorly constrained. Additionally, we test the effect of only removing $H$, with the IR still being constrained by the other bands. We also find no significant difference in stellar masses, with the median of the difference being 0 and the average -0.08 (M$_{\rm \star, no\,H}$ being slightly smaller). Overall, we find that not removing photometric bands outside Ly$\alpha$ does not lead to a significant overestimation of stellar masses for our sources. However, including nebular lines may still be important, particularly if we look at other parameters (e.g. ages), as there may be some systematics, particularly for the faintest sources with the highest EWs. This will be addressed in a forthcoming paper with an SED-fitting code that models nebular emission \citep[{\sc CIGALE}, ][]{Noll2009,Boquien2019}.

For our $z\sim2-6$ LAEs, the optical bands are essential to fit the rest-frame UV continuum, IRAC filters can constrain fluxes redward of D$_{4000}$ and the FIR measurements provide upper constraints in the dust emission, which can improve the SFR estimates. We note that, as explained in $\S$\ref{sec:sample_AGNs}, while we exclude sources with evidence of AGN activity when computing median properties of the sample, we still obtain SED fits (without using any AGN SED model) for those sources.

In Fig. \ref{fig:SED}, we show observed and intrinsic SED fits and photometric measurements/upper limits for two LAEs. The SEDs were purposely chosen to show two very distinct galaxies within the SC4K sample: one with a very blue and steep UV continuum slope, with low stellar mass that dominate the sample and one with a more red continuum, more massive and with higher dust extinction which is much more rare in the sample of LAEs. While the latter is not well representative of a typical LAE, it is still important to show that LAEs can span a large variety of physical properties. This LAE is detected in two {\it Herschel} bands, which shows that FIR can be important to constrain the SED fits and derive properties of high redshift LAEs.

\subsubsection{Number of derived SEDs} \label{sec:number_seds}

Although all LAEs are by definition detected in the MB/NB where they were selected \citep{Sobral2018}, a small fraction of our LAEs have few to no detections in other photometric bands. For such cases, SED-fitting may fail. Out of the 3590 non-AGN LAEs, we obtain reliable SEDs for 3377 (94\%, see Table \ref{tab:overview}). The catalogue that we release with this paper (see \S\ref{sec:catalogue}) has an SED flag which marks unreliable SEDs. The AGN flag indicates AGN LAEs (\S\ref{sec:sample_AGNs}, \citealt{Calhau2019}), and we reiterate that while we compute parameters for these sources, the SED-derived parameters are not reliable and are not included in any median property estimation done in this work.

%
%
\begin{figure*}
  \centering
  \begin{tabular}{cc}
  \includegraphics[width=0.49\textwidth]{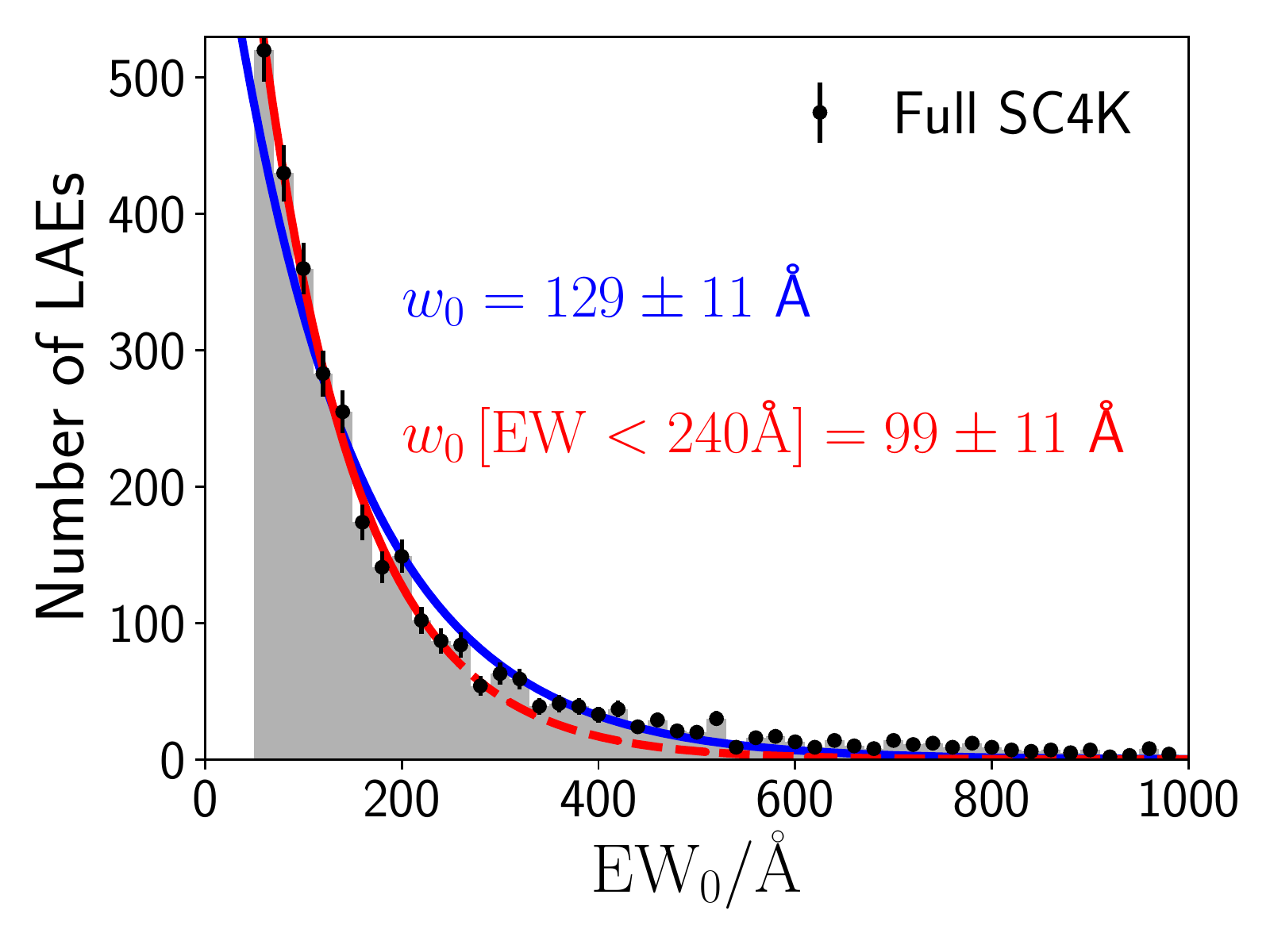}
  \includegraphics[width=0.49\textwidth]{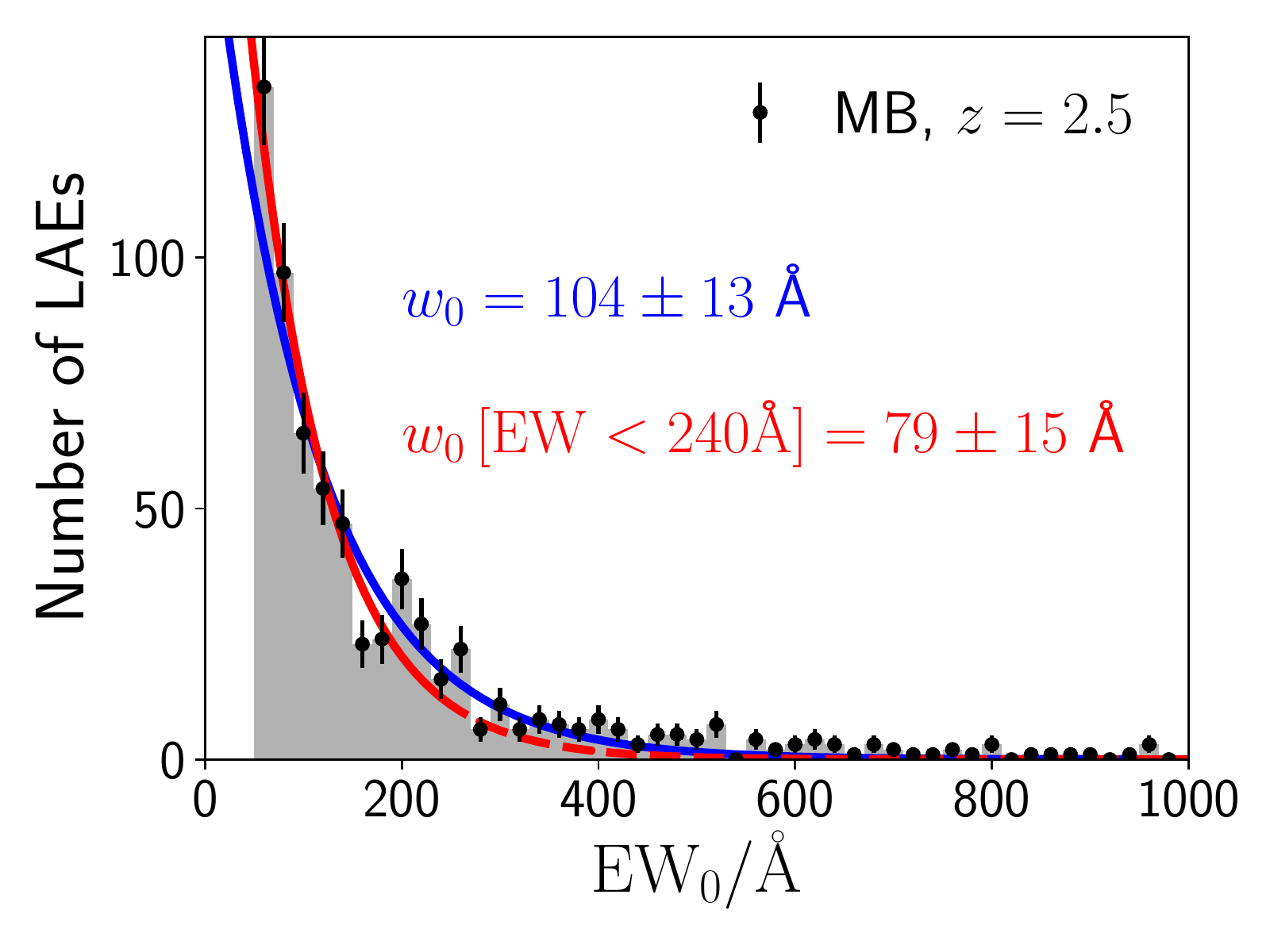}
  \end{tabular}
   \caption{{\it Left:} EW$_0$ distribution of the full SC4K sample of LAEs. We fit an exponential function of the form {\rm $N=N_0$ exp(-EW$_0/w_0$)}, and derive the parameter $w_0$. Fit derived with the distribution of EW$_0$ (EW$_0<240$\,\AA) is shown in red (blue).  {\it Right:} Same but for an individual filter (IA427) with LAEs at $z=2.5$.}
  \label{fig:w0}
\end{figure*}

\section{The properties of LAEs} \label{sec:methods}

In this section, we present our methodology and computations to derive galaxy properties for individual LAEs, using our full photometric measurements and SED fits from {\sc MAGPHYS}. EW$_0$ and L$_{\rm Ly\alpha}$ of all LAEs in the SC4K sample have been derived and published in \cite{Sobral2018}.

\subsection{Ly$\alpha$ luminosity (L$_{\rm Ly\alpha}$)} \label{subsec:EW_Lya}

L$_{\rm Ly\alpha}$ is calculated from the Ly$\alpha$ line flux ($f_{\rm Ly\alpha}$):

\begin{equation}
{\rm L}_{\rm Ly\alpha} [\rm erg\,s^{-1}]= 4\pi f_{\rm Ly\alpha}{\rm D_L^2}(\it z) 
\end{equation}
where D$_{\rm L}(z)$ is the luminosity distance at the redshift of each source, computed from the redshifted Ly$\alpha$ at the effective wavelength of the detection NB/MB. In Fig. \ref{fig:histograms1} (left) we show the L$_{\rm Ly\alpha}$ distribution of our LAEs, spanning a wide range of luminosities L$_{\rm Ly\alpha}=10^{42-44}$ erg s$^{-1}$.

\subsection{Ly$\alpha$ rest-frame equivalent width (EW$_0$)}

The observed EW (EW$_{\rm obs}$) of an emission line is the ratio between the flux of the line and the continuum flux density and can be calculated as:

\begin{equation}
\rm EW_{\rm obs} [\text{\normalfont\AA}]= \Delta\lambda_{1}\frac{\it f_{1}-f_{2}}{\it f_{2}-f_{2}(\Delta\lambda_{1}/\Delta\lambda_{2})},
\end{equation}
where $\Delta\lambda_{1}$ is the FWHM of the NB or MB, $\Delta\lambda_{2}$ the excess broad band filter \citep{Sobral2018}, $f_{1}$ is the flux density measured in the NB or MB and $f_{2}$ is the flux density computed from two adjacent BB filters, which avoids assumptions of the slope of the continuum \citep[for details see][]{Sobral2018}. The rest-frame EW (EW$_{0}$) is calculated as:
\begin{equation}
{\rm EW}_{\rm 0} [\text{\normalfont\AA}]= \frac{\rm EW_{\rm obs}}{1+z},
\end{equation}
where $z$ is the redshift of Ly$\alpha$ at the effective wavelength of the NB or MB \citep{Sobral2018}. We provide the median EW$_0$ for different redshifts and for the full SC4K sample in Table \ref{tab:overview}.

%
%
\begin{figure*}
  \centering
  \includegraphics[width=\textwidth]{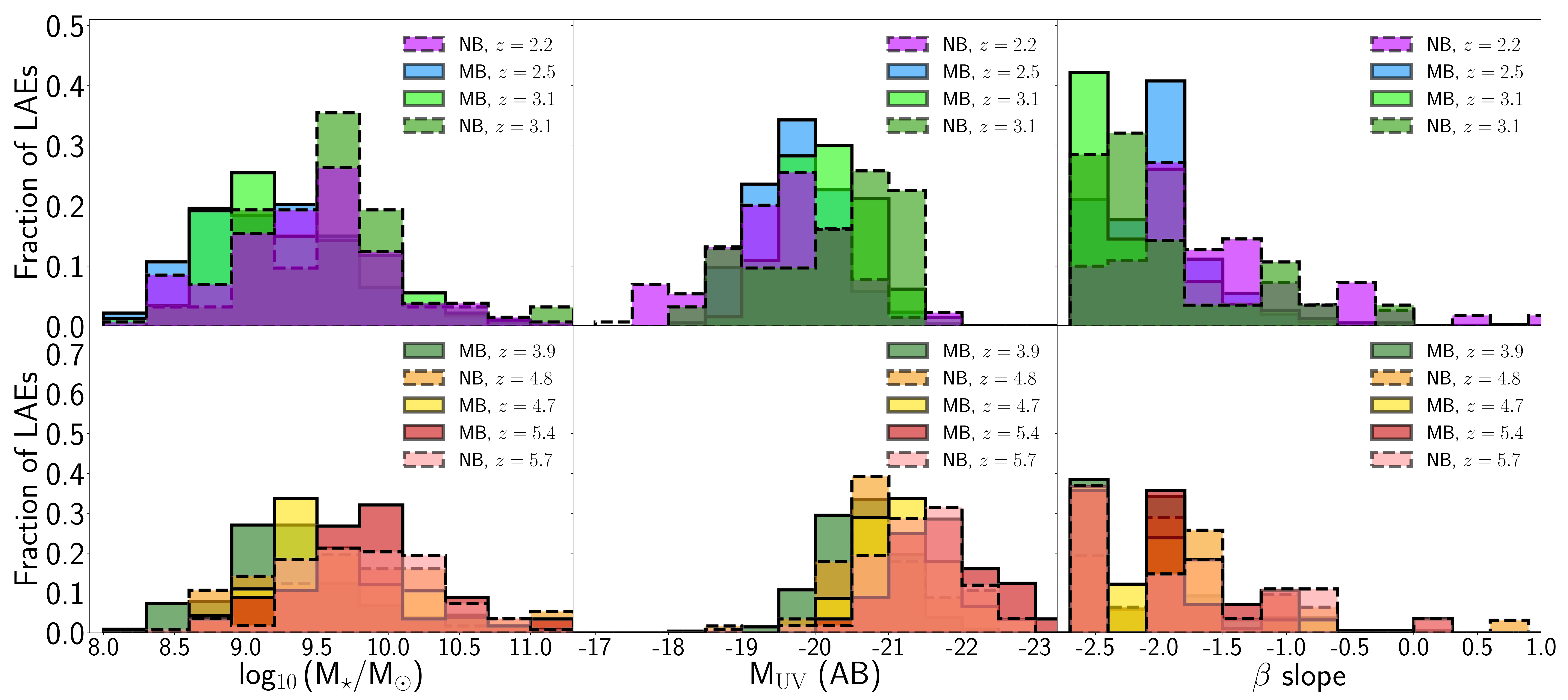}
  \caption{Distribution of properties derived from the SED fitting ({\sc MAGPHYS}, see $\S$\ref{subsec:methods_SED}). We show the stellar mass, M$_\star$ (left), rest-frame UV luminosity, M$_{\rm UV}$ (middle) and rest-frame UV slope, $\beta$ (right). Top panels show the $z\leq3.1$ sample and the bottom panels show the higher redshift LAEs. AGNs have been removed.}
  \label{fig:histograms2}
\end{figure*}

\subsubsection{EW$_0$ scale length ($w_0$)} \label{sec:w0}

An exponential fit of the form {\rm $N=N_0$ exp(-EW$_0/w_0$)} has been widely used to describe Ly$\alpha$ EW$_0$ distributions \citep[e.g.][]{Gronwall2007,Hashimoto2017,Wold2017}, with the rate of decay being determined by the scale length parameter $w_0$. With our sample of LAEs, we analyse EW$_0$ distributions in multiple well defined redshift ranges between $z\sim2$ and $z\sim6$. To estimate $w_0$, we define bins of 20\,\AA\, and fit the exponential function to the observed distribution (see Figure \ref{fig:w0}), taking into account Poissonian errors. Bins with less than two sources are excluded from the fits. To account for bin width choice, we add 10\,\AA\,(half the bin width) in quadrature to the errors of $w_0$. We also explore how an EW$_0$ upper cut affects $w_0$ as it removes sources with extreme (and more uncertain) EWs. We apply a cut of EW$_0=240$\,\AA, the theoretical limit of EW$_0$ powered by Population II star formation \citep[e.g.][]{Charlot1993} and the value which has been extensively used in Ly$\alpha$ emission studies to identify ``extreme" EW galaxies \citep[e.g.][]{Cantalupo2012,Marino2018}. We compute $\chi^2_{\rm red}$ by comparing the best exponential fit to the histogram of observed counts and their associated Poisson errors.
 
Additionally, we fully explore how the errors on EW$_0$ influence the measurement of $w_0$ by using an MCMC approach. For each iteration, we perturb the EW$_0$ of each LAE in that specific sample within their asymmetric error bars (assuming a double normal probability distribution function centred at each EW$_0$ and with FWHM equal to the errors derived from photometry; \citealt{Sobral2018}). We impose a hard lower limit equal to the detection threshold ($50$\,\AA\, for MBs, $25$\,\AA\, for NBs except for NB392 which has a lower limit of $5$\,\AA; see $\S$\ref{subsec:sample}) and an upper limit of 1000\,\AA, with any source outside these values not being included in a specific realisation. With the perturbed EW$_0$, we construct the histogram of the current iteration, using bins of 20\,\AA. We fit an exponential to the generated histogram bins, taking into account the associated Poissonian error ($\sqrt{N}$) of each bin. We iterate this process 200 times, and the final $w_0$ is the median value of all fits with error up (down) being the 84th (16th) percentile of all fits. In addition, to account for the uncertainty introduced by the bin width choice, we also add 10\,\AA\, in quadrature to the errors of $w_0$. We also apply the MCMC approach with a cut of EW$_0=240$\,\AA. For the MCMC approach, where EW$_0$ are perturbed, $\chi^2_{\rm red}$ is computed by comparing the best fit to the median histogram of all iterations and its Poisson errors.

In Table \ref{tab:w0}, we show the inferred $w_0$ values (including perturbed estimates) for different redshift ranges and filter combinations. 

Furthermore, it is important to establish how the EW$_0$ distribution depends on M$_{\rm UV}$ and M$_\star$. To understand this dependence, we measure $w_0$ in three M$_{\rm UV}$ and M$_\star$ ranges and show our measurements in Table \ref{tab:w0}. For the faintest and the lowest mass ranges, we are significantly incomplete to the low EW$_0$ end of the EW distribution, resulting in a peak at $\sim100$\,\AA. Thus, we only fit EW$_0>100$\,\AA\, to accurately estimate the exponential decay of the distribution for these two cases.

\subsection{Rest-frame UV luminosity (M$_{\rm UV}$)} \label{subsec:muv}

The UV luminosity of a galaxy is associated with continuum emission from massive stars and traces SFR in the past 100 Myr \citep[e.g.][]{Boselli2001,Salim2009}. A priori, sources selected by their strong Ly$\alpha$ emission could be expected to have strong M$_{\rm UV}$ as both trace recent star formation (neglecting AGN contribution), although Ly$\alpha$ can trace slightly more recent star formation because stars dominating the ionising photon budget have lifetimes of $\sim10$ Myr. However, as shown by e.g. \cite{Matthee2017spectra} and \cite{Sobral2018} more factors come into play as Ly$\alpha$ and M$_{\rm UV}$ do not necessarily correlate with each other, due to e.g. highly ISM dependent f$_{\rm esc, Ly\alpha}$ (which can result in most Ly$\alpha$ emission being absorbed by dust particles or scattered off neutral hydrogen) or an ionising efficiency which is evolving with redshift.

We compute M$_{\rm UV}$ by integrating the best-fit SEDs at rest-frame $\lambda_0=1400-1600$\,\AA. We show the M$_{\rm UV}$ histogram distribution in Figure \ref{fig:histograms2} (centre). Due to the magnitude limits, at higher redshift we are only sensitive to more luminous M$_{\rm UV}$ sources. We detect SC4K LAEs as bright as M$_{\rm UV}=-23$ and as faint as M$_{\rm UV}=-17$.

\subsection{UV continuum slope ($\beta$)} \label{subsec:beta}

The slope of the UV continuum can be parametrised in the form $f_\lambda \propto \lambda^\beta$ \citep[e.g.][]{Meurer1999}. The slope $\beta$ is sensitive to the age, metallicity and dust content of a galaxy. \cite{Bruzual2003} models used by {\sc MAGPHYS} have a hard limit to how negative (blue) $\beta$ can be ($\beta=-2.44$), a natural consequence of an upper limit in the IMF. While $\beta$ may be intrinsically even bluer for more ``extreme" stellar populations, in this study, we do not explore those.

We measure $\beta$ directly from the best-fit as the slope of the continuum at rest-frame $\lambda_0=1300-2100$\,\AA. We apply a  conservative approach and only use $\beta$ measurements from sources with at least two detections in this wavelength range. This ensures the $\beta$ slope is directly constrained and not a direct consequence of assumed SED templates. As expected, due to an increasing luminosity distance, combined with rest-frame $\lambda_0=1300-2100$\,\AA\,moving into IR wavelengths, there are fewer $\beta$ measurements at higher redshift. In addition, we also compute $\beta$ by fitting a power-law ($\beta_{\rm pl}$) to the photometric measurements \citep[similar to e.g.][]{Bouwens2014}, with no SED fitting assumptions. We fit $\beta_{\rm pl}$ in the range $\lambda_0=1400-2100$\,\AA, which is smaller than the range used to compute $\beta$ from the SED fit to avoid broad band filters at $\sim1300$\,\AA, which can be contaminated by the Ly$\alpha$ break. Only sources with at least three 3$\sigma$ detections in that range are considered for the power-law measurement. For the full SC4K, we measure a median $\beta_{\rm pl}=-1.8^{+0.8}_{-0.7}$, which is redder (+0.3) than $\beta$ from the SED fit (0.2 when only considering sources with $\beta_{\rm pl}$ measurements), but still within the error bars. Overall, $\beta$ is better constrained through SED fitting as it uses a prior - the SED models included in {\sc MAGPHYS}. Furthermore, the SED models take into account $\sim30$ filters over the full UV-FIR wavelength range, preventing it from being as sensitive to individual filter measurements in the smaller $\lambda_0=1400-2100$\,\AA\,range. Thus, throughout this paper, we use $\beta$ computed from the SED fits.

We show the histogram distributions of $\beta$ in Fig. \ref{fig:histograms2} (right). LAEs tend to be very blue across all redshift ranges (median $\beta=-2.1^{+0.5}_{-0.4}$, Table \ref{tab:overview}). LAEs at $z=2.2$ are found to have the reddest $\beta$ slopes, albeit still very blue and comparable to the Lyman Break Galaxy (LBG) population (see further discussion in \ref{sec:MUV_beta}). We note, nonetheless, that the $z=2.2$ sample has some key differences compared to other LAEs in SC4K sample, as it selects LAEs down to 5\,\AA\, EW$_0$ in addition to reaching the faintest L$_{\rm Ly\alpha}$. This allows redder sources to be picked up, while the much higher EW$_0$ LAEs tend to have much bluer $\beta$ slopes.

\subsection{Stellar Mass (M$_\star$)} \label{subsec:mstar}

The total mass of stars in a galaxy (stellar mass, M$_\star$) is a fundamental galaxy property which is a reflection of its star formation history. We use M$_\star$ derived from the likelihood parameter distribution from {\sc MAGPHYS} modelling.

We show the histogram distribution of M$_\star$ in our sample in Figure \ref{fig:histograms2} (left). Most LAEs (88\%) have stellar masses $<10^{10}$\,M$_\odot$, although it is important to stress there are some more massive galaxies, which shows a significant diversity. We observe a slight shift to higher masses as we move to higher redshifts (see also Table \ref{tab:overview}) but this is a natural consequence of only being sensitive to intrinsically more luminous galaxies at higher redshift. We find that typical LAEs are low stellar mass galaxies, with the median of the SC4K sample of LAEs being M$_\star=10^{9.3^{+0.6}_{-0.5}}$\,M$_\odot$.

%
%
\begin{figure*}
  \centering
  \includegraphics[width=\textwidth]{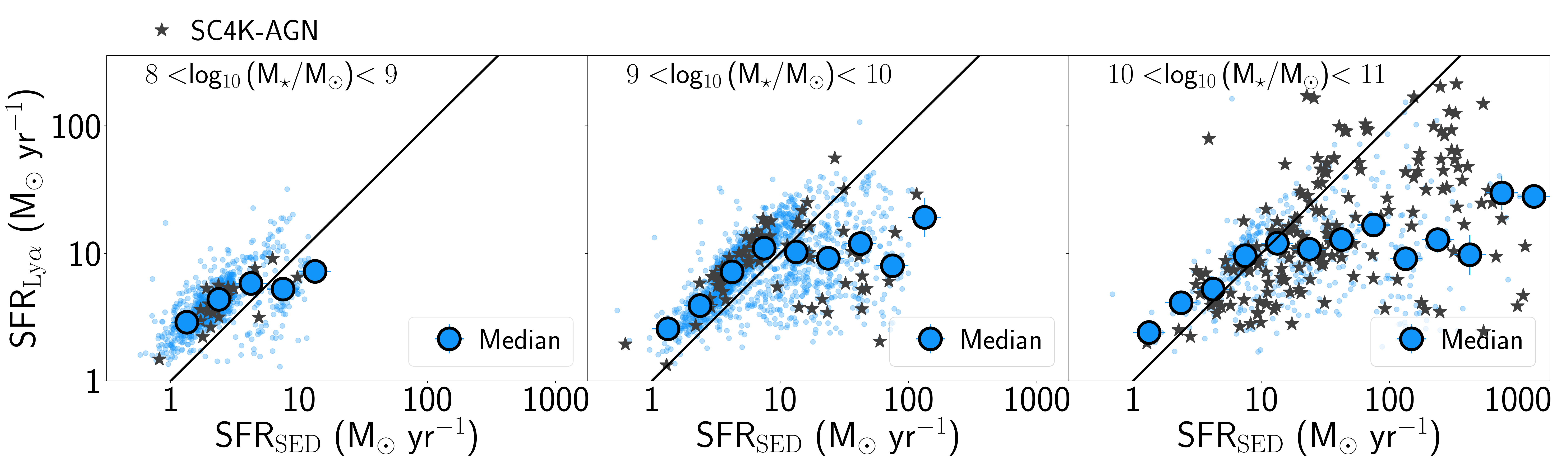}
  \caption{Emission line-based SFR vs SED-fitting SFR for the full sample of LAEs at different stellar masses. Blue circles are the median bin and individual points are plotted as scatter in the background. The black line is the 1-to-1 ratio. There is a small systematic offset at M$_\star=10^{8-9}$\,M$_\odot$ yr$^{-1}$ and for SFR$_{\rm SED}=1-10$\,M$_\odot$ yr$^{-1}$ for all stellar mass ranges. For higher stellar masses and SFRs, there is a more significant difference between the two methods, with the emission line-based approach predicting lower SFRs. This is a likely consequence of Ly$\alpha$ not being sensitive to obscured regions in very massive galaxies, thus not being sensitive to their full contribution. Additionally, we plot AGN LAEs with black stars (purely stellar+dust SED-fitting with no AGN models) to show they are typically measured as having high stellar masses when blindly running SED codes with no AGN models in AGN samples.}
  \label{fig:SFR_SM19vsSFR_SED_Mstar}
\end{figure*}

\subsection{Star Formation Rates (SFRs)} \label{subsec:methods_SFR}

\subsubsection{Emission line-based SFRs with Ly$\alpha$} \label{subsec:methods_SFR_Lya}

We estimate the SFR directly from ${\rm L}_{\rm Ly\alpha}$ and EW$_0$, using the recipe from \cite{SobralMatthee2019} which has calibrated EW$_0$ as a good empirical indicator of f$_{\rm esc, Ly\alpha}$. With a measurement of f$_{\rm esc, Ly\alpha}$, ${\rm L}_{\rm Ly\alpha}$ can be converted to dust-corrected H$\alpha$ luminosity assuming case-B recombination \citep{Brocklehurst1971} and transformed into SFR following \cite{Kennicutt1998}. For a Chabrier IMF ($0.1-100$\,M$_\odot$) and assuming f$_{\rm esc,LyC}=0$, ${\rm L}_{\rm Ly\alpha}$ in erg s$^{-1}$ and EW$_0$ in \AA, the SFR thus becomes \cite{SobralMatthee2019}:

\begin{equation}\label{eq:SFR}
{\rm SFR_{Ly\alpha}\,[{\rm M_{\odot}\,yr^{-1}}]}=\frac{{\rm L}_{\rm Ly\alpha}\times4.4\times10^{-42}}{\rm0.042\,{\rm EW}_{0}},
\end{equation}

For EW$_{0}>210$\,\AA, following \cite{SobralMatthee2019}, we set f$_{\rm esc, Ly\alpha}=1$ which corresponds to SFR\,$[{\rm M_{\odot}\,yr^{-1}}]=4.98\times10^{-43}\times{\rm L}_{\rm Ly\alpha}$, with ${\rm L}_{\rm Ly\alpha}$ in erg s$^{-1}$. This SFR is calibrated with dust-corrected H$\alpha$ luminosities and thus should be interpreted as dust-corrected SFR. We show the SFR distribution in Figure \ref{fig:histograms1} (right). As the SFR is derived from ${\rm L}_{\rm Ly\alpha}$, it is limited by the same detection limits, which causes a shift to higher SFR with increasing redshift. We measure SFRs in the range $\sim1-300\,{\rm M_{\odot}\,yr^{-1}}$, and measure a median SFR$_{\rm Ly\alpha}=5.9^{+6.3}_{-2.6}$ for SC4K LAEs (see Table \ref{tab:overview}).

\subsubsection{SED-derived SFRs} \label{subsec:methods_SFR_SED}

As previously stated, {\sc MAGPHYS} uses a bayesian approach to estimate the best likehood SFR, comparing model SEDs (generated using some assumptions, see \S\ref{subsec:methods_SED}) with observed photometry. Due to our FIR measurements being mostly upper limits for $>99\%$ of SC4K LAEs, it is not possible to directly measure the amount of SFR that is obscured by dust and the optical thickness of dust from IR-FIR. As such, the amount of dust and SFR is inferred from the UV-optical slope. We measure SFRs in the range  $\sim0.1-3000\,{\rm M_{\odot}\,yr^{-1}}$, and measure a median SFR$_{\rm SED}=4.4^{+10.5}_{-2.4}$\,M$_\odot$ yr$^{-1}$ for SC4K LAEs (Table \ref{tab:overview}).

\subsubsection{SFR$_{\rm Ly\alpha}$ vs SFR$_{\rm SED}$} \label{sec:SFR_comparison}

In this work, we estimate SFRs of individual LAEs using two approaches: emission line-based with Ly$\alpha$ (SFR$_{\rm Ly\alpha}$, \S\ref{subsec:methods_SFR_Lya}) and from SED-fitting (SFR$_{\rm SED}$, \S\ref{subsec:methods_SFR_SED}). These two approaches are independent as SFR$_{\rm Ly\alpha}$ is derived directly from two properties of the Ly$\alpha$ emission-line (luminosity and EW$_0$), while SFR$_{\rm SED}$ is obtained with {\sc MAGPHYS} by removing the filter contaminated by Ly$\alpha$ and using up to $\approx30$ photometric data-points from the rest-frame UV to the rest-frame FIR.

In Fig. \ref{fig:SFR_SM19vsSFR_SED_Mstar} we show a comparison between SFR$_{\rm Ly\alpha}$ and SFR$_{\rm  SED}$ at different mass ranges. We measure a small systematic offset at M$_\star=10^{8-9}$\,M$_\odot$ and SFR$_{\rm SED}=1-10$\,M$_\odot$ yr$^{-1}$ for all stellar mass ranges, with the emission line-based approach predicting slightly higher SFRs. As Ly$\alpha$ traces more recent star-formation than the UV-continuum, the higher predicted SFRs could be explained by on-going bursts of star-formation, which lead to slightly higher SFR$_{\rm Ly\alpha}$. Only for SFRs which are measured to be high from SED (SFR$_{\rm SED}>10$\,M$_\odot$ yr$^{-1}$) there is a significant difference, with SFR$_{\rm Ly\alpha}$ being lower and its median maxing at $\approx10$\,M$_\odot$ yr$^{-1}$. Such SFR ranges are typically only seen in more massive ranges (M$_\star=10^{9-11}$\,M$_\odot$), which are thus more susceptible to have underestimated SFRs from Ly$\alpha$. This is in line with what could be expected for very massive galaxies as Ly$\alpha$ will only be able to measure the contribution in regions of the galaxy which are actively star-forming and unobscured, leading to underestimated SFRs in these regimes. Nevertheless, it is remarkable that two largely independent methods obtain such similar results. For the global populations of SC4K LAEs, these two methods also retrieve very similar SFRs of $5.9^{+6.3}_{-2.6}$ and $4.4^{+10.5}_{-2.4}$\,M$_\odot$ yr$^{-1}$ for the emission line-based and SED-based, respectively. Additionally, in the Appendix (Fig. \ref{fig:SFR_SM19vsSFR_SED}), we show SFR$_{\rm Ly\alpha}$ vs SFR$_{\rm SED}$ at different redshift ranges. Both approaches predict very similar SFRs at all redshifts, outside the differences at aforementioned ranges as the emission line-based approach cannot reach such SFR ranges.

Furthermore, in a recent study by \cite{Calhau2019}, the SFR of the SC4K sample is derived through the stacking of radio imaging in the 3GHz band. For the stacking procedure, individual sources with direct detections are removed as these are likely AGN. They find median SFR$_{\rm radio}=5.1^{+1.3}_{-1.2}$\,M$_\odot$ yr$^{-1}$ from the $z\sim2-6$ stack, which is in very good agreement with emission line-based and SED-based SFR estimates of the sample.

%
%
\begin{figure*}
\centering
\includegraphics[width=0.98\textwidth]{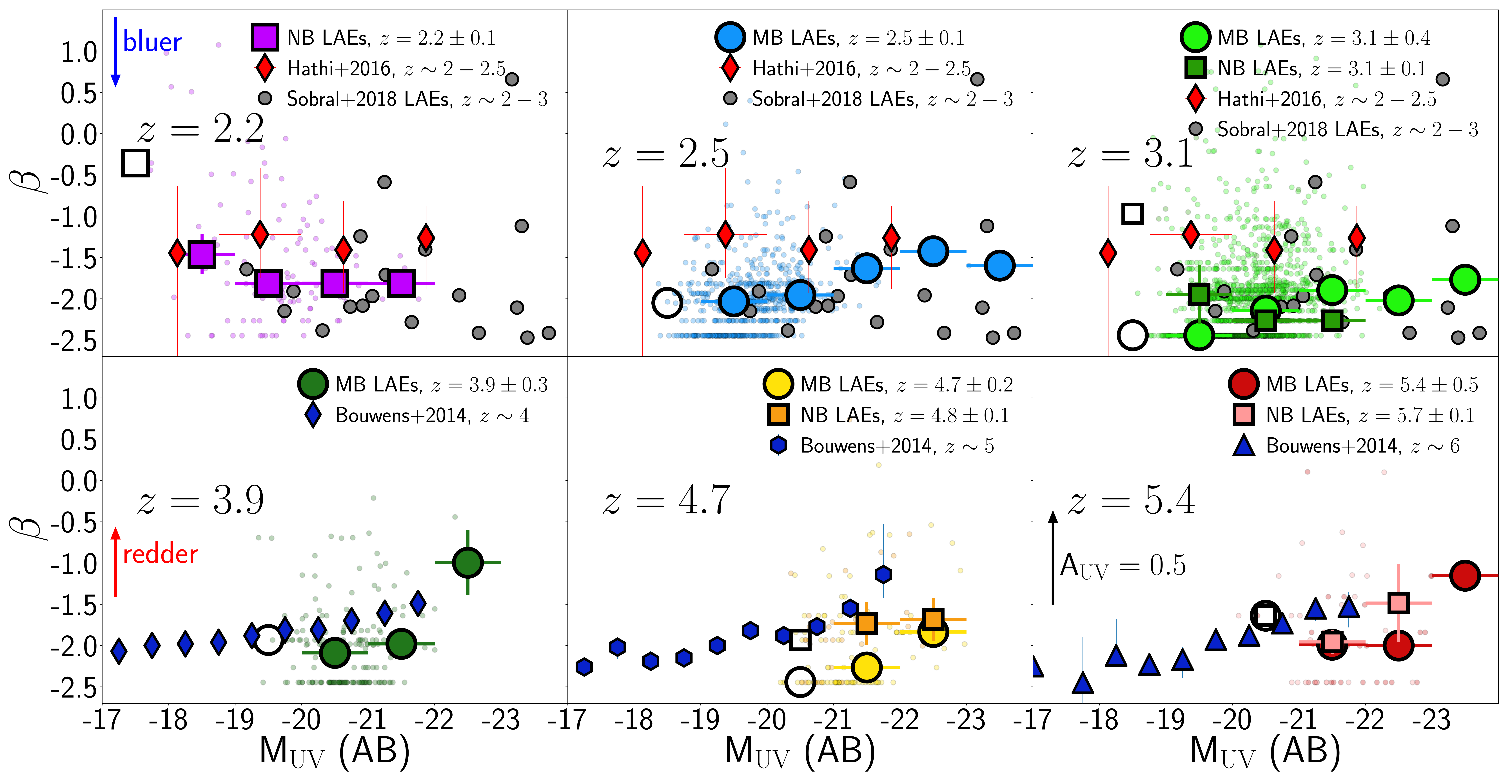}
\caption{UV-continuum slope $\beta$ (measured from SED fitting, see $\S$\ref{subsec:beta}) vs UV luminosity M$_{\rm UV}$ (derived by integrating the SED fits at $\sim$1500\,\AA, see \S\ref{subsec:muv}). Each panel contains LAEs from different redshift intervals (from left to right $z=2.2, 2.5, 3.1, 3.9, 4.7, 5.4$). The median $\beta$ of each M$_{\rm UV}$ bin of LAEs selected through medium (narrow) band filters is shown as filled coloured circles (squares) with the individual points being plotted as scatter in the background. Unfilled markers are likely biased bins, as discussed in $\S$\ref{sec:MUV_beta}. The clustering of points at $\beta=-2.44$ is a physically imposed model limitation as $\beta$ can not become bluer without increasing the upper mass of the IMF to unreasonable values. For comparison we add measurements from LAEs at $z\sim2-3$ \citep{Sobral2018Spectra} and UV-continuum selected samples at $z\sim2-2.5$ \citep[][]{Hathi2016} and $z\sim4$, $z\sim5$ and $z\sim6$ \citep[][]{Bouwens2014}. The black arrow is the size in $\beta$ of A$_{\rm UV}=0.5$ (A$_{\rm UV}=4.43 + 1.99\beta$, \citealt{Meurer1999}). We find the median $\beta$ in LAEs to be as blue or bluer than UV-selected samples at the same M$_{\rm UV}$ for all redshifts.}
  \label{fig:beta_MUV}
\end{figure*}

\subsection{Catalogue of SC4K LAE properties} \label{sec:catalogue}

With this paper, we make public a catalogue with multiple measurements for individual LAEs in the SC4K sample. For each LAE we provide R.A., Dec, L$_{\rm Ly\alpha}$, EW$_0$, X-ray and radio Flags (as given by \citealt{Sobral2018}) and updated X-ray and radio Flags (as given by \citealt{Calhau2019}), M$_\star$, $\beta$, M$_{\rm UV}$, SFR$_{\rm Ly\alpha}$ and SFR$_{\rm SED}$, with associated errors. We also provide our photometric measurements in Jansky for the 34 filters used in this work and a boolean SED flag which indicates unreliable SEDs. For LAEs with True SED flag, we set all SED-derived properties to -99. We provide the catalogue of SC4K LAEs in electronic format in Appendix \ref{ap:cat}.

\section{Results and Discussion} \label{sec:results}

\subsection{M$_{\rm UV}-\beta$ relation for LAEs and its evolution}  \label{sec:MUV_beta}

The UV rest-frame luminosity (M$_{\rm UV}$) and the UV $\beta$ slope follow a tight correlation in UV-continuum selected samples \citep[e.g.][]{Bouwens2014}, with faint M$_{\rm UV}$ galaxies being typically bluer (more negative $\beta$). We measure how these two parameters are correlated for LAEs, whether they follow a similar M$_{\rm UV}$-$\beta$ relation as UV-continuum selected samples, and whether the relation evolves.

In Fig. \ref{fig:beta_MUV}, we show the relation between M$_{\rm UV}$ (\S\ref{subsec:muv}) and $\beta$ (\S\ref{subsec:beta}) for 6 redshift intervals ($z=2.2, 2.5, 3.1, 3.9, 4.7, 5.4$). We note that at very faint M$_{\rm UV}$ we are biased towards redder sources. This is a consequence of redder sources being easier to detect in the optical filters, while sources with a very steep continuum slope will fall below our detection limits, particularly faint M$_{\rm UV}$ sources. As such, in Fig. \ref{fig:beta_MUV}, we show the faintest M$_{\rm UV}$ bin as unfilled.

%
%
\begin{figure}
  \centering
  \includegraphics[width=0.47\textwidth]{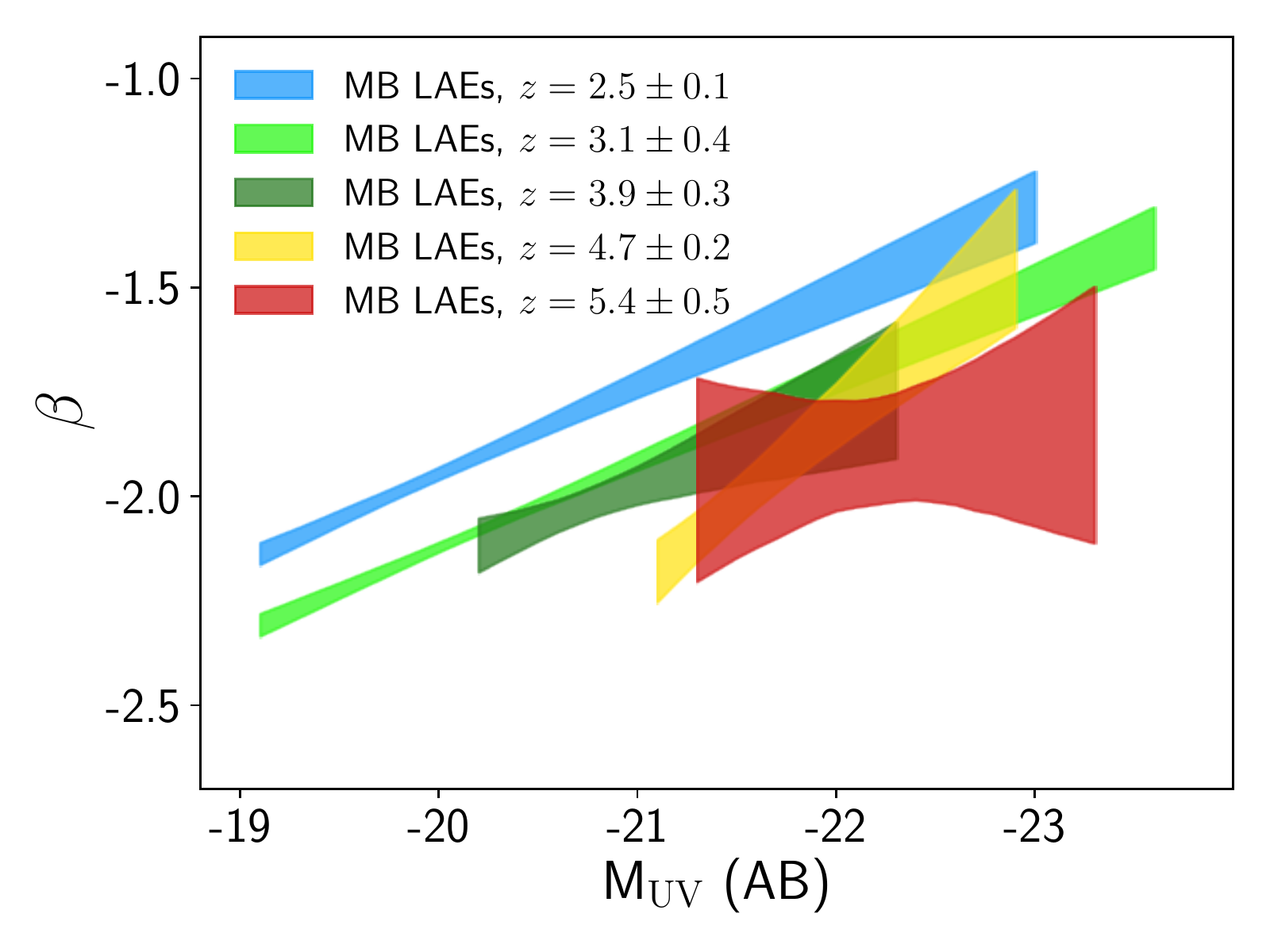}
  \caption{The evolution of the M$_{\rm UV}$-$\beta$ relation for LAEs. Shaded regions are the 1\,$\sigma$ intervals obtained by bootstrapping the individual measurements for which we are not significantly biased (see $\S$\ref{sec:MUV_beta}). $\beta$ increases with M$_{\rm UV}$ and this relation shifts down to smaller $\beta$ as we move to higher redshifts. Most of this trend seems to be captured by a decrease in the normalisation of the relation, but we also find some evidence of the relation steepening.}
  \label{fig:beta_contour_mb}
\end{figure}

%
%
\begin{figure*}
\centering
\includegraphics[width=0.75\textwidth]{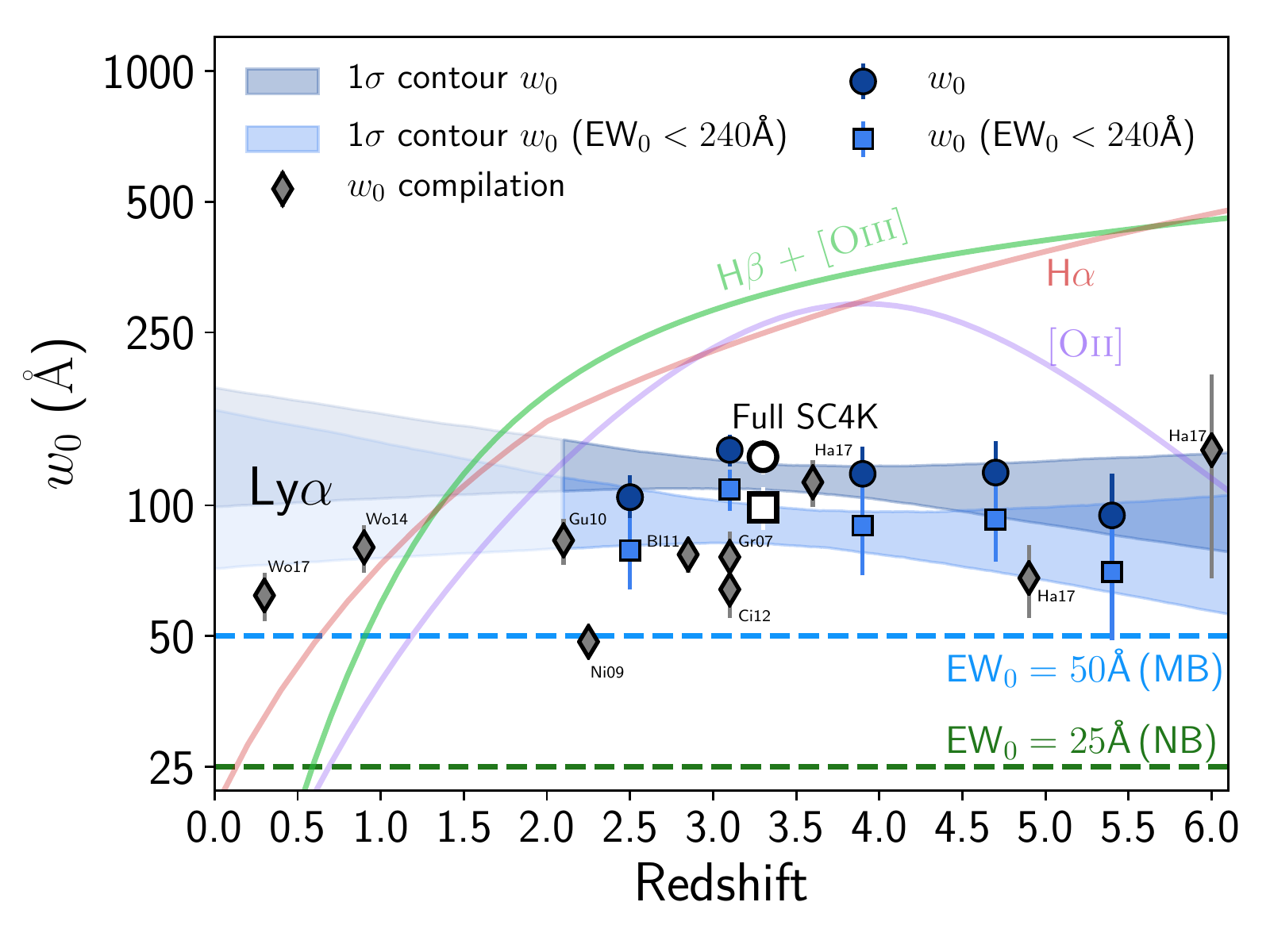}
\caption{Global Ly$\alpha$ $w_0$ evolution with redshift. Best $w_0$ estimates are shown as blue circles (squares) for the full range of EW$_0$ (EW$_0<240$\,\AA). Blue contours are estimated by perturbing the $w_0$ bins within error bars (see $\S$\ref{sec:w0} for details). We find evidence for little to no evolution of $w_0$. The white points show Ly$\alpha$ $w_0$ of the full SC4K sample. We present a compilation of Ly$\alpha$ $w_{0}$ from $z=0.3$ to $z\sim6$ \citep[][]{Gronwall2007, Nilsson2009, Guaita2010, Blanc2011, Ciardullo2012, Wold2014,Wold2017, Hashimoto2017}. In addition, we show the {\sc [Oii]} (H$\beta$ + {\sc [Oiii]}) rest-frame equivalent widths of emitters selected by these lines \citep{Khostovan2016} as purple (green) fits and H$\alpha$ EW$_0$ \citep{Faisst2016,Matthee2017LyC} as red. Overall, the consensus of all data points is that there is no significant Ly$\alpha$ $w_0$ evolution with redshift despite the strong increase in the typical EW$_0$ of non-resonant lines for a wider population of SFGs.}
  \label{fig:w0_evolution}
\end{figure*}

LAEs are found to be consistently bluer than UV-selected samples \citep{Bouwens2014,Hathi2016} at similar redshifts (up to $\sim1$ dex bluer), regardless of being NB or MB-selected, at all redshifts studied \citep[see also][]{Hashimoto2017}. Our results are consistent with $z\sim2-3$ LAEs measurements from \cite{Sobral2018Spectra}. Additionally, we measure an increase of $\beta$ with M$_{\rm UV}$ ($\sim0.5$ dex per $\Delta$M$_{\rm UV}=2$), indicating that brighter M$_{\rm UV}$ LAEs are redder at all redshift ranges, even though LAEs are typically bluer. This tight correlation between M$_{\rm UV}$ and $\beta$ is very similar to the one observed in LBG populations, implying an important overlap between the populations and also an important diversity within the LAE population.

In Fig. \ref{fig:beta_contour_mb}, we show the 1\,$\sigma$ contours for the M$_{\rm UV}$ vs $\beta$ distribution. We compute the 1\,$\sigma$ contours by bootstrapping our individual data points. We choose a random subset of 50\% of the data points, determine the best fit, iterate the process 1000 times and define the 1\,$\sigma$ contours as the 16th and 84th percentiles of all fits. As previously mentioned, faint M$_{\rm UV}$ bins will be biased towards redder sources, which are easier to detect in the continuum. As such, we apply a M$_{\rm UV}$ cut to our fits, equal to the lower limit of the faintest filled M$_{\rm UV}$ bin (Fig. \ref{fig:beta_MUV}).

Overall we find a M$_{\rm UV}$-$\beta$ relation for LAEs, which is qualitatively very similar to the one observed in UV-selected samples. As it can be seen in Fig. \ref{fig:beta_contour_mb}, the normalisation of the M$_{\rm UV}$-$\beta$ relation slowly moves to bluer $\beta$ with increasing redshift for LAEs, and there is also tentative evidence for the relation to become steeper at higher redshift. This can also be seen in Fig. \ref{fig:beta_MUV}, where the lowest redshift LAEs show a much flatter relation, while at higher redshift the relation seems to be steeper. These results might be explained due to a consistent average decrease in dust content and metallicity even within LAEs from low to high redshift.

\subsection{Implications of M$_{\rm UV}-\beta$ relation for LAEs}  \label{sec:interpUVB}

The UV continuum $\beta$ slope can be an indicator of the dust attenuation of a galaxy as well as the age and metallicity of its stellar population, but because it is sensitive to all these effects, it can also be very complicated to interpret \citep[see e.g.][]{Popping2017}. As shown by \cite{Bouwens2012} (see Fig. 13 therein), a negative offset of $\sim0.5-1$\,dex in $\beta$ should be dominated by a change in dust, albeit age and metallicity can also significantly steepen $\beta$, with a hotter population of stars. This suggests that LAEs are a subset of the SFG population which is very young and likely more metal-poor, with significant contribution from O and B stars which make the UV continuum steeper.

In LBGs, $\beta$ has been shown to depend on the UV luminosity, with a similar slope independent of redshift \citep[e.g.][]{Bouwens2012,Bouwens2014}. The normalisation of the relation is shifted to bluer $\beta$ as we move to higher redshifts which can be explained by a lower dust content/lower dust extinction in galaxies at higher redshift \citep[e.g.][]{Finkelstein2012}. As shown in Fig. \ref{fig:beta_contour_mb}, LAEs have a very similar behaviour to LBG galaxies: $\beta$ is tightly correlated with M$_{\rm UV}$, with brighter M$_{\rm UV}$ galaxies being redder and the normalisation of this slope shifting to lower $\beta$ with increasing redshift, which can be explained by a lower dust content at higher redshift even for LAEs. Similar observations of the M$_{\rm UV}$-$\beta$ trend and the $\beta$ evolution with redshift have been shown by \cite{Hashimoto2017}. The work presented by \cite{Hashimoto2017} reaches fainter M$_{\rm UV}$ than the work presented here and thus provides a consistent view of UV properties in LAEs from a complementary work using a different selection method (integral field spectroscopy with MUSE).

\subsection{Ly$\alpha$ EW$_{0}$ and $w_0$: evolution for LAEs?} \label{sec:ew}

EW$_{0}$ is an indicator of the strength of an emission line relatively to the continuum. As such, it holds important information about a galaxy, with high EW$_{0}$ being associated with young stellar ages, low metallicities and top-heavy IMFs \citep{Schaerer2003,Raiter2010}. We use our sample of LAEs at well-defined redshift ranges to probe for redshift evolution of EW$_{0}$.

%
%
\begin{figure*}
\centering
\includegraphics[width=0.47\textwidth]{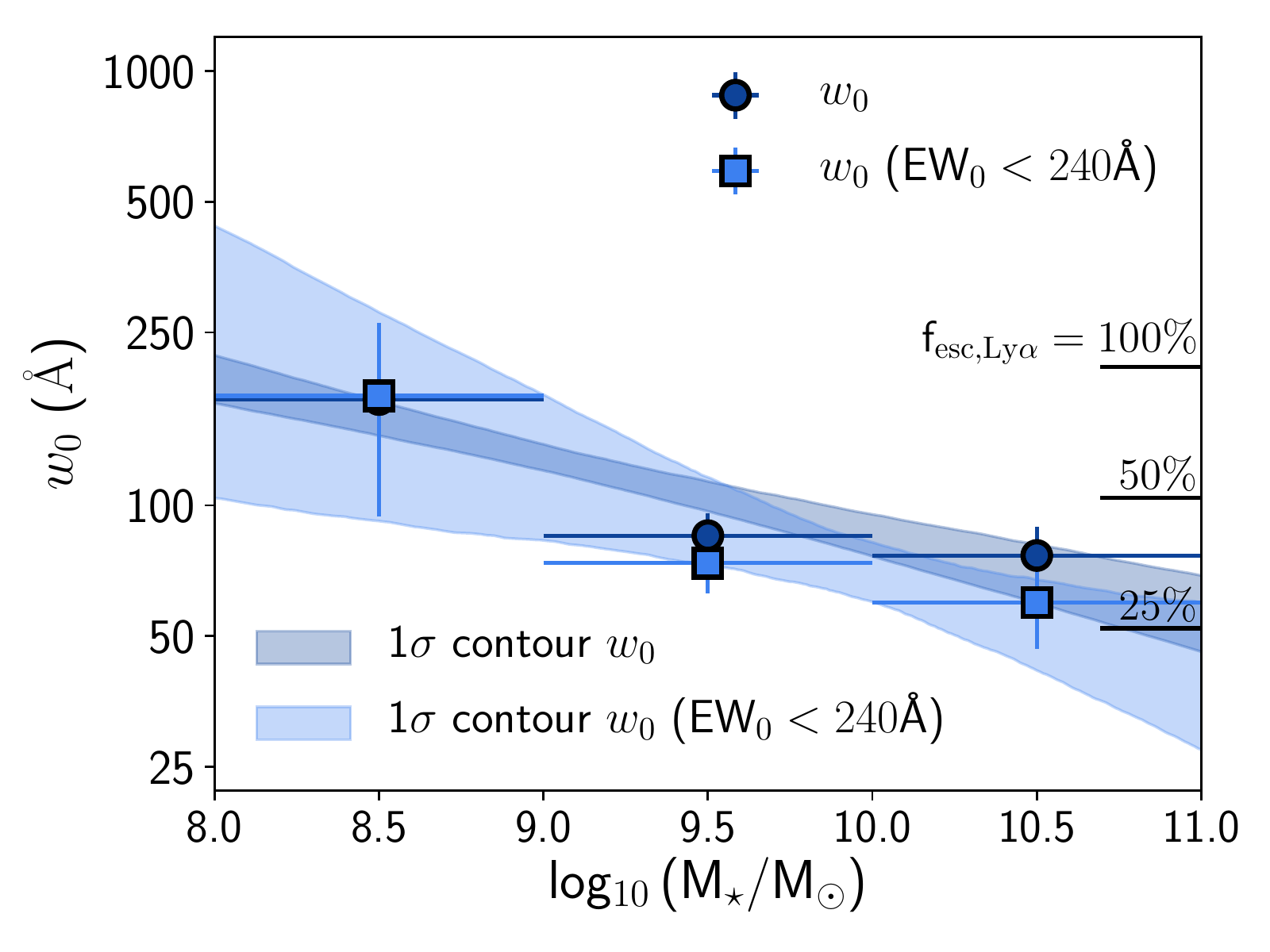}
\includegraphics[width=0.47\textwidth]{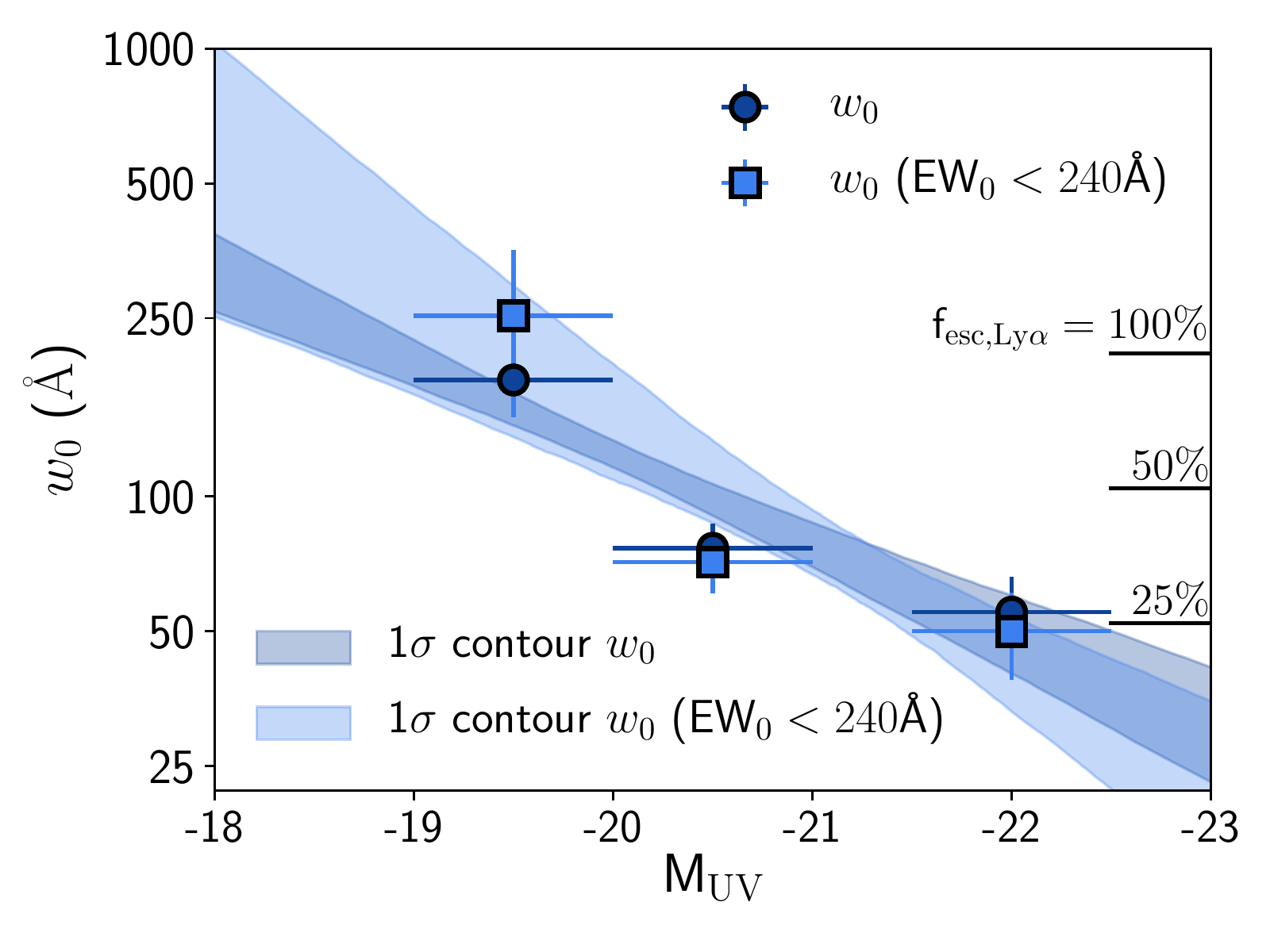}
\caption{The Ly$\alpha$ $w_{0}$ dependence on M$_\star$ and M$_{\rm UV}$. Best $w_0$ estimates are shown as blue circles (squares) for the full range of EW$_0$ (EW$_0<240$\,\AA). A label with f$_{\rm esc, Ly\alpha}$ ($=0.048 w_0$; \citealt{SobralMatthee2019}) is added for a potential physical interpretation of results. {\it Left:} Ly$\alpha$ $w_{0}$ is anti-correlated with stellar mass, such that the most massive LAEs have the lowest $w_0$ and likely the lowest f$_{\rm esc, Ly\alpha}$. {\it Right:} Ly$\alpha$ $w_{0}$ is also anti-correlated with UV luminosity, with the faintest UV LAEs having the highest Ly$\alpha$ $w_{0}$.}
  \label{fig:w0_dep_MUV_Mstar}
\end{figure*}
We find the median Ly$\alpha$ EW$_{0}$ of SC4K LAEs to remain constant at $\sim140$\,{\AA} with redshift, both in MB and NB-selected samples (median EW$_0=138^{+284}_{-70}$\,\AA). We show the little to no evolution of median EW$_{0}$ in Fig. \ref{fig:EW_evolution}. For individual filters, we detect a tentative higher than average EW$_0$ at $z\sim5.7-5.8$, which could be caused by the small sample size or higher contamination fraction, although we highlight the large error bars.

The calculated median Ly$\alpha$ EW$_{0}$ can be very sensitive to selection effects, and it is possible that the non-evolution we measure is a consequence of the relatively high EW$_{0}>50$\,{\AA} cut applied in SC4K. In order to further tackle this, we also investigate the evolution of the scale parameter $w_0$ ($\S$\ref{sec:w0}). $w_0$ has been extensively probed in the literature \citep[see e.g.][]{Ciardullo2012,Hashimoto2017}, particularly because the exponential decay of the EW$_0$ distribution should be less affected by observational EW$_0$ cuts.

Our results are presented in Fig. \ref{fig:w0_evolution}. We find no statistically significant evolution of the Ly$\alpha$ $w_0$ with redshift. Generally, $w_0$ is slightly higher when determining it without any upper constraints on the Ly$\alpha$ EW$_0$, and lower if we restrict its calculation to LAEs with EW$_0$, but no significant evolution is seen when using a single self-consistent method. We therefore conclude that both the observed median Ly$\alpha$ EW$_0$ and the distributions of Ly$\alpha$ $w_0$ for LAEs are not changing significantly from $z\sim2$ to $z\sim6$. A non-evolution of $w_0$ suggests there is no significant evolution in the typical or average properties of sources selected as LAEs across cosmic time. These include their typical metallicities and dust properties, but also perhaps more importantly their Ly$\alpha$ escape fraction, f$_{\rm esc, Ly\alpha}$. As shown by \cite{SobralMatthee2019}, the observed Ly$\alpha$ EW$_0$ can be used to estimate f$_{\rm esc, Ly\alpha}$. The non-evolution of Ly$\alpha$ EW$_0$ and $w_0$ across time implies non-evolving f$_{\rm esc, Ly\alpha}$ for LAEs. For SC4K LAEs, we infer a constant f$_{\rm esc, Ly\alpha}$ of $\approx0.6-0.7$ across cosmic time ($\approx0.5-0.6$ when applying the EW$_0>240$\,\AA\, cut). These median f$_{\rm esc, Ly\alpha}$ values are consistent with those derived using radio SFRs for SC4K Ly$\alpha$ emitters \citep[$0.7\pm0.2$, see][]{Calhau2019}.

However, it should be noted that different redshifts do not necessarily probe the same M$_{\rm UV}$ ranges (Fig. \ref{fig:histograms2}, middle panel), which should be considered when discussing $w_0$ evolution with redshift, particularly as $w_0$ depends on M$_{\rm UV}$ (see \S\ref{subsec:w0_dependence}). We attempt to explore potential bias effects by computing $w_0$ with a consistent M$_{\rm UV}$ cut. For the full SC4K sample, we compute $w_0$ for -22<M$_{\rm UV}$<-19 which is a M$_{\rm UV}$ range probed by all redshifts (see middle panel of Fig. \ref{fig:histograms2}). For this cut, a flat relation (non-evolution) is still observed within 0.9\,$\sigma$ for EW$_0<240$\,\AA\, and 1.8\,$\sigma$ for the full EW$_0$ range. The different M$_{\rm UV}$ ranges probed by different selection filters/redshifts do not seem to be sufficient to explain the non-evolution of $w_0$ with redshift, which is likely a characteristic of the LAE population itself.

\subsubsection{Comparison with other studies}

In order to compare our results with other studies across different redshifts, in Fig. \ref{fig:w0_evolution} we show a compilation of Ly$\alpha$ $w_{0}$ in samples of LAEs, from $z\sim0$ to $z\sim6$ \citep[][]{Gronwall2007, Nilsson2009, Guaita2010, Blanc2011, Ciardullo2012, Wold2014,Wold2017, Hashimoto2017}. Our results agree well with \cite{Hashimoto2017}, \cite{Guaita2010} and \cite{Blanc2011}. Furthermore, our extrapolation of $w_0$ to low redshift is consistent with the results from \cite{Wold2014,Wold2017}.

Our measurements reveal higher values than those by \cite{Nilsson2009}, \cite{Gronwall2007} and \cite{Ciardullo2012}, all at intermediate redshifts ($z=2.25-3.1$) and with selections that go to much lower EWs. We note however that the $w_0$ measured by \cite{Nilsson2009} is below our MB detection threshold and that our blind selection of LAEs is not sensitive to the lowest EW$_0$, as highlighted in Fig. \ref{fig:w0_evolution}. Our LAE selection of high EW LAEs is much more similar to blind surveys done with MUSE \citep{Hashimoto2017}, but SC4K allow the selection and study of much higher luminosity LAEs. Furthermore, we note that our $w_0$ measurements shift to smaller values when the EW$_0<240$\,\AA\,cut is applied, becoming even more similar to the measurements reported in the literature. 

While there are observed variations due to different sample selections which contribute to the scatter (Fig. \ref{fig:w0_evolution}), overall we conclude that there is no clear evolution of the Ly$\alpha$ EW$_{0}$ and $w_0$ for LAEs when taking into account all measurements. Such parameters remaining constant for LAEs contrasts with measurements from other non-resonant emission lines for the general star-forming population, which are found to increase significantly with redshift. In order to provide a rough comparison, in Fig. \ref{fig:w0_evolution} we also show the redshift evolution of the rest-frame EW of line-emitters, including {\sc [Oii]} and H$\beta$ + {\sc [Oiii]} emitters \citep{Khostovan2016} and H$\alpha$ EW$_0$ \citep{Sobral2014}. While at $z\sim0$ those non-resonant rest-frame optical emission lines have typical EW$_{0}<25$\,{\AA}, by $z\sim2$ they already exceed Ly$\alpha$ EW$_{0}$. This reveals a very significant evolution of the typical stellar populations of the general population of SFGs, while those selected to be LAEs have high Ly$\alpha$ EW$_{0}$ at all cosmic times. Since LAEs have typically high EWs in their rest-frame optical lines, it is very likely that we are seeing star-forming galaxies becoming, on average, LAEs, towards $z\sim6$. Such possibility would easily explain the rise in the global Ly$\alpha$/UV luminosity densities \citep[see full discussion and implications in][]{Sobral2018}.

\subsubsection{The $w_{0}$ and f$_{\rm esc, Ly\alpha}$ dependence on M$_\star$ and M$_{\rm UV}$} \label{subsec:w0_dependence}

LAEs seem to show no evolution in their typical Ly$\alpha$ $w_0$ across cosmic time. However, one could expect that LAEs with different physical properties may show different $w_0$, particularly as a consequence of different Ly$\alpha$ escape fractions \citep[see e.g.][]{Matthee2016,Oyarzun2017,SobralMatthee2019}.

We start by investigating how Ly$\alpha$ $w_0$ may depend on the stellar mass of LAEs. The results are presented on the left panel of Fig. \ref{fig:w0_dep_MUV_Mstar}, where we show the results when restricting the measurements to EW$_0<240$\,{\AA} and when using full samples. We find an anti-correlation between Ly$\alpha$ $w_0$ and stellar mass, with the least massive LAEs having $w_0\approx180$\,\AA\, and the most massive having $w_0\approx70$\,\AA. By using \cite{SobralMatthee2019}, this could be seen as a significant difference in the typical f$_{\rm esc, Ly\alpha}$ which would decline from $\approx90$\% for $\rm M\sim10^{8.5}$\,M$_{\odot}$ LAEs to f$_{\rm esc, Ly\alpha}\approx30\%$ for $\rm M\sim10^{10.5}$\,M$_{\odot}$ LAEs. This trend is very similar to those found by \cite{Matthee2016} for a general population of H$\alpha$ emitters with much higher SFRs and lower f$_{\rm esc, Ly\alpha}$ than our LAEs and by \cite{Oyarzun2017}.

In Fig. \ref{fig:w0_dep_MUV_Mstar} (right panel) we also show how Ly$\alpha$ $w_0$ is clearly anti-correlated with M$_{\rm UV}$. Our results show that UV luminous LAEs in our sample (M$_{\rm UV}\approx-21.5$) have Ly$\alpha$ $w_0\approx50$\,{\AA}, which rises with declining UV luminosity to $w_0\approx180$\,{\AA} for M$_{\rm UV}\approx-19.5$ LAEs. This implies that the UV faintest sources have the highest f$_{\rm esc, Ly\alpha}$ \citep[][]{SobralMatthee2019} of around $\approx85$\%, while the most UV luminous LAEs have f$_{\rm esc, Ly\alpha}\approx20-30$\%. Our results are in good agreement with \cite{Oyarzun2017} and reveal that even though LAEs have high Ly$\alpha$ $w_0$ across cosmic time, the population still shows important trends with stellar mass and rest-frame UV luminosity.

\subsubsection{LAEs with extreme EW$_0$}

The nature of LAEs with extremely high EW$_0$ and the processes behind the creation of such extreme lines are still a relatively unexplored topic despite a range of discoveries \citep[e.g.][]{Cantalupo2012,Kashikawa2012,Hashimoto2017,Maseda2018}. Typical internal star formation processes should not be enough to power EW$_0>240$\,\AA\, in Ly$\alpha$ \citep[][]{Schaerer2003,Raiter2010}, but studies like \cite{Cantalupo2012} suggest that such extreme objects which have been found could be explained by fluorescent ``illumination" from e.g. a nearby quasar \citep[see also][]{Rosdahl2012,Yajima2012}. Additionally, an extreme $z=6.5$ LAE with EW$_0=436$\,\AA\, is reported in \cite{Kashikawa2012}, with the authors arguing that such a high EW$_0$ requires a very young, massive and metal-poor stellar population, or even Population III stars.

The large volume covered by SC4K ($\sim10^8$\,Mpc$^3$) and the sensitivity to the highest EWs provides a unique opportunity to identify and quantify the number density of extremely high EW LAEs. In order to do so in a conservative way, rather than simply selecting sources with Ly$\alpha$ EW$_0$ higher than 240\,{\AA}, we take the photometric errors fully into account, and we use the 3\,$\sigma$ errors. In practice, we look for LAEs within SC4K which satisfy EW$_0>240$\,{\AA} at a 3\,$\sigma$ level \footnote{EW$_0-3\Delta$EW$_0>240$\,\AA} and for which we have no evidence of AGN activity. We find a total of 45 ``extreme" non-AGN LAEs in $\sim61.5\times10^6$\,Mpc$^3$ and we investigate how these are distributed across redshift. The results are shown in Table \ref{tab:extreme_EW}, where we use Poisson errors. Most of the extreme LAEs are found at $z\sim2-3$. Furthermore, by taking into account the volumes surveyed, we find that the number density of extreme LAEs within SC4K rises, from $(0.12\pm0.08)\times10^{-6}$\,Mpc$^{-3}$ at $z\sim5.4$ to $(1.50\pm0.61)\times10^{-6}$\,Mpc$^{-3}$ at $z\sim2.5$, although such increase should be treated with caution, as the higher redshift sample does not reach very faint M$_{\rm UV}$ ($>-20$) ranges. Overall, we find a number density of $(0.73\pm0.11)\times10^{-6}$\,Mpc$^{-3}$ at $z\sim2-6$, revealing that these sources are exceptionally rare. At 1\,$\sigma$ confidence level, we find 318 LAEs with EW$_0>240$\,\AA, resulting in a number density of $(5.17\pm0.29)\times10^{-6}$\,Mpc$^{-3}$. Spectroscopic follow-up observations are required to further understand their nature. We find our 45 ``extreme" sources to be a diverse population, as they are found at all Ly$\alpha$ luminosities and stellar masses, but preferentially at faint UV luminosities which is a consequence of high EW + ``random" Ly$\alpha$ luminosities. They typically have blue UV $\beta$ slopes but some reach redder values ($\beta\sim-1.2$). We do not observe a spatial correlation between ``extreme" LAEs and AGN, which we would expect if the high EWs in this sample of LAEs were generated by fluorescent ``illumination".

Through a narrow band filter search, \cite{Cantalupo2012} targeted a field centred in a hyper luminous quasar and identified 18 LAEs at $z=2.4$ in a comoving volume of 5500\,Mpc$^3$. Stacking of these sources results in EW$_0>800$\,\AA\, (1\,$\sigma$), which cannot be explained by typical star-formation processes. This implies a higher number density of extreme LAEs than the conservative number density we report in this paper, although this can be easily explained by \cite{Cantalupo2012} specifically targeting a quasar field.

In a more comparable blank search, using deep MUSE data, \cite{Hashimoto2017} selected 6 LAEs with EW$_0>240$\,\AA\, at a 1\,$\sigma$ level (zero at 3\,$\sigma$) in $9.31\times10^4$\,Mpc$^3$ \citep{Drake2017} at $z\sim2-6$. This results in a number density of $\sim6\times10^{-5}$\,Mpc$^{-3}$, suggesting these ``extreme" LAEs may be even more common at fainter luminosities than those in the SC4K sample.

%
%
\begin{table}
\begin{center}
\caption{Number count and number density of LAEs with EW$_0>240$\,{\AA} at a 3\,$\sigma$ level, for different redshift intervals, using comoving volumes from \citet{Sobral2018}. Errors are Poissonian. We find very low number densities of extreme LAEs, but these increase with decreasing redshift.} \label{tab:extreme_EW}
\begin{tabular}{c | cc}
\hline
\multicolumn{1}{c|}{Redshift interval} &
\multicolumn{1}{c|}{N} &
\multicolumn{1}{c|}{$\Phi$} \\
& (\# LAEs) & ($10^{-6}$\,Mpc$^{-3}$)\\
\hline
MB, $z=2.5\pm0.1$ & 6 ($\pm2$) & $1.50\pm0.61$ \\
MB, $z=3.1\pm0.4$ & 15 ($\pm4$) & $0.82\pm0.21$ \\
MB, $z=3.9\pm0.3$ & 4  ($\pm2$) & $0.40\pm0.20$ \\
MB, $z=4.7\pm0.2$ & 2  ($\pm1$) & $0.17\pm0.12$ \\
MB, $z=5.4\pm0.5$ & 2  ($\pm1$) & $0.12\pm0.08$ \\
\hline
Full sample & 45  ($\pm7$) & $0.73\pm0.11$ \\
\hline
\end{tabular}
\end{center}
\end{table}

%
%
\begin{figure*}
\centering
\includegraphics[width=0.9\textwidth]{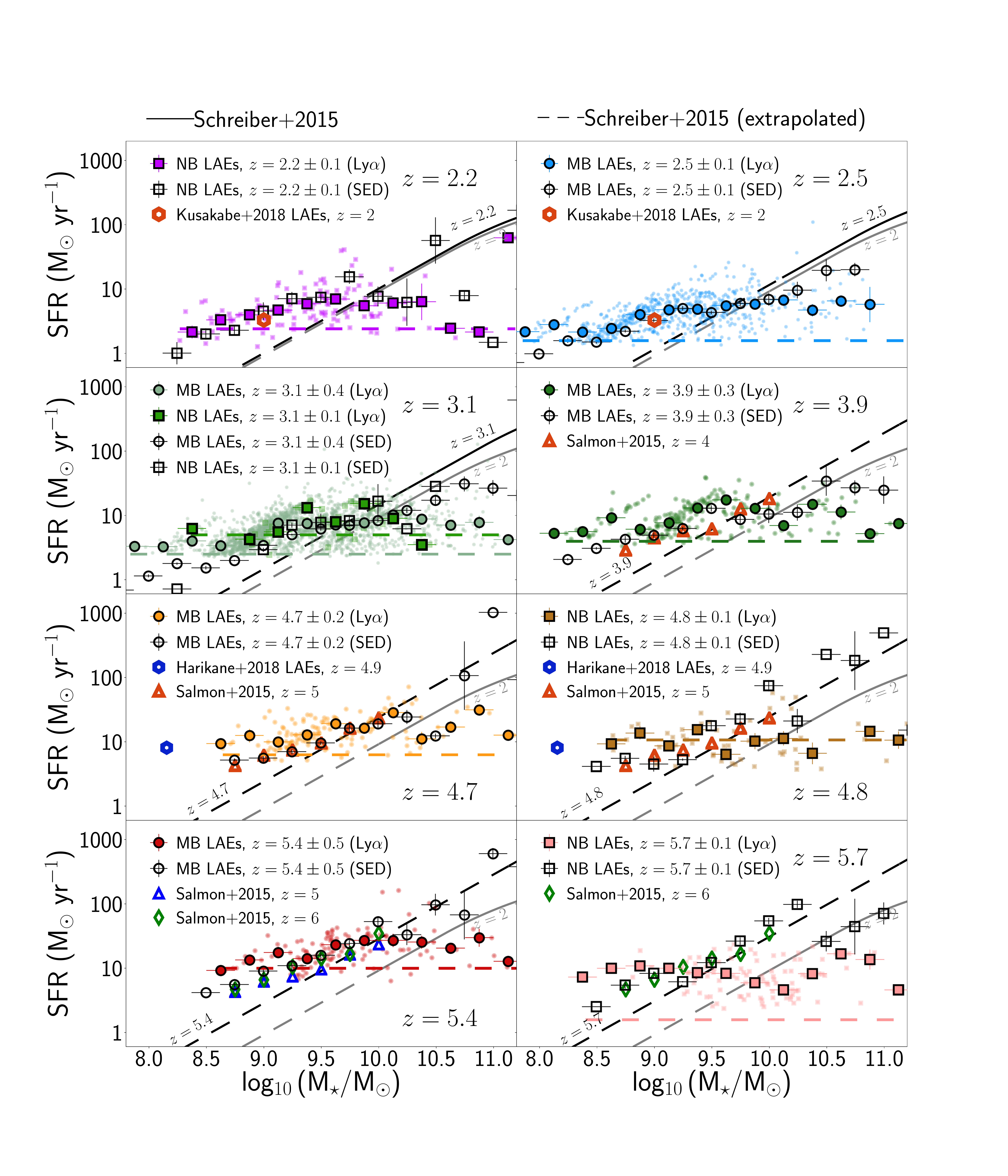}
\caption{SFR (derived from Ly$\alpha$ and EW$_0$ and derived from SED fits, see \S\ref{subsec:methods_SFR}) vs M$_\star$ (derived from SED fits, see \S\ref{subsec:methods_SED}). Each panel contains LAEs from different redshift intervals (from left to right $z=2.2, 2.5, 3.1, 3.9, 4.7, 5.4$). The median SFR$_{\rm Ly\alpha}$ of each M$_\star$ bin for LAE selected through medium (narrow) band filters is shown as filled coloured circles (squares) with the individual points being plotted as scatter in the background. The median SFR$_{\rm SED}$ of each M$_\star$ bin for LAE selected through medium (narrow) band filters is shown as open circles (squares). The dotted horizontal line is the average SFR depth, computed from the flux depth and average EW$_0$ of the sample. The continuous black lines are the best-fit relations from \citet{Schreiber2015} computed for the redshift of each panel and converted from Salpeter to Chabrier. These relations are shown as dashed lines for the mass ranges where they were extrapolated.} \label{fig:sfr_sm19_mstar}
\end{figure*}

%
%
\begin{figure*}
\centering
\begin{tabular}{cc}
\includegraphics[width=0.49\textwidth]{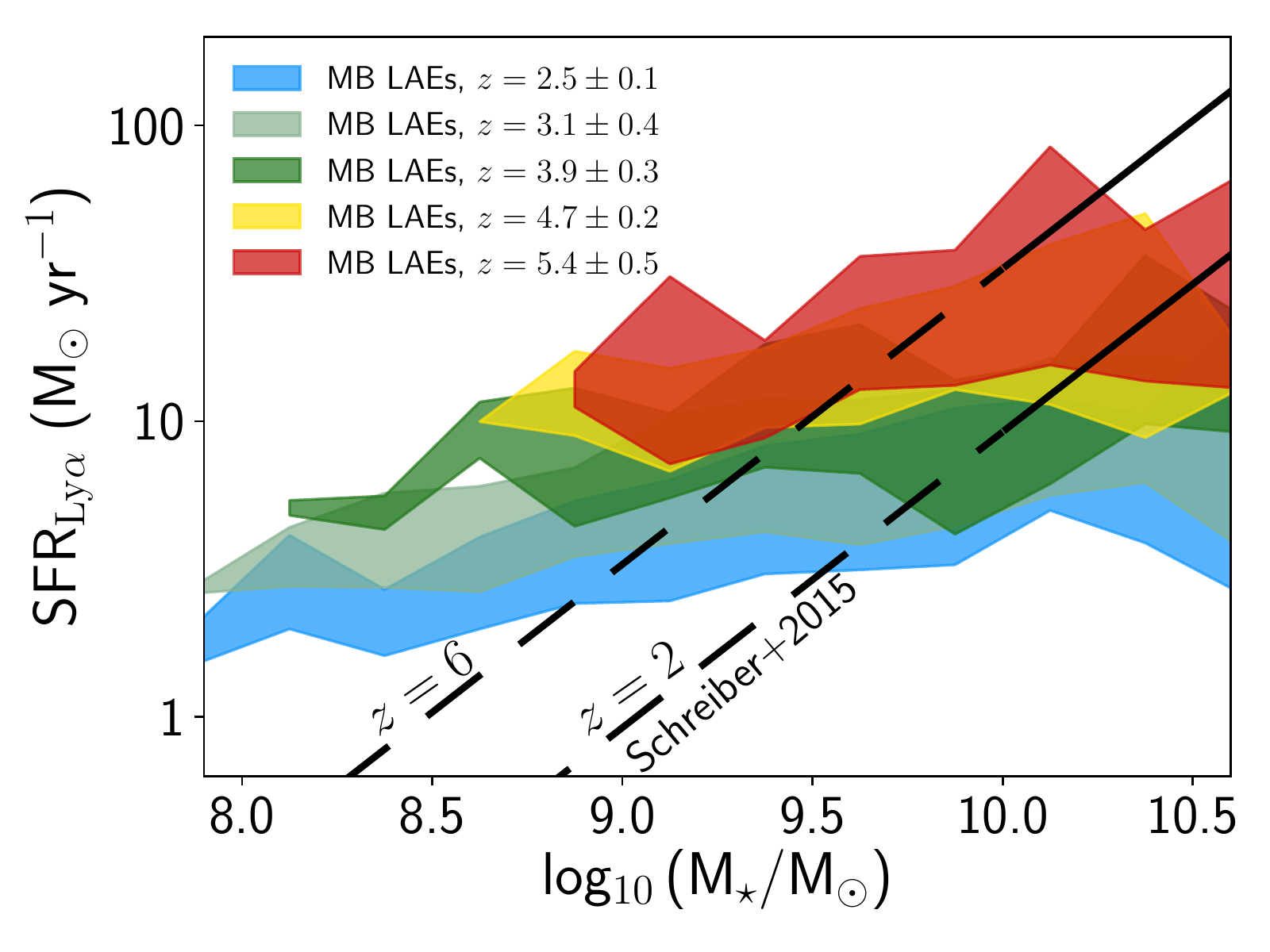} 
\includegraphics[width=0.49\textwidth]{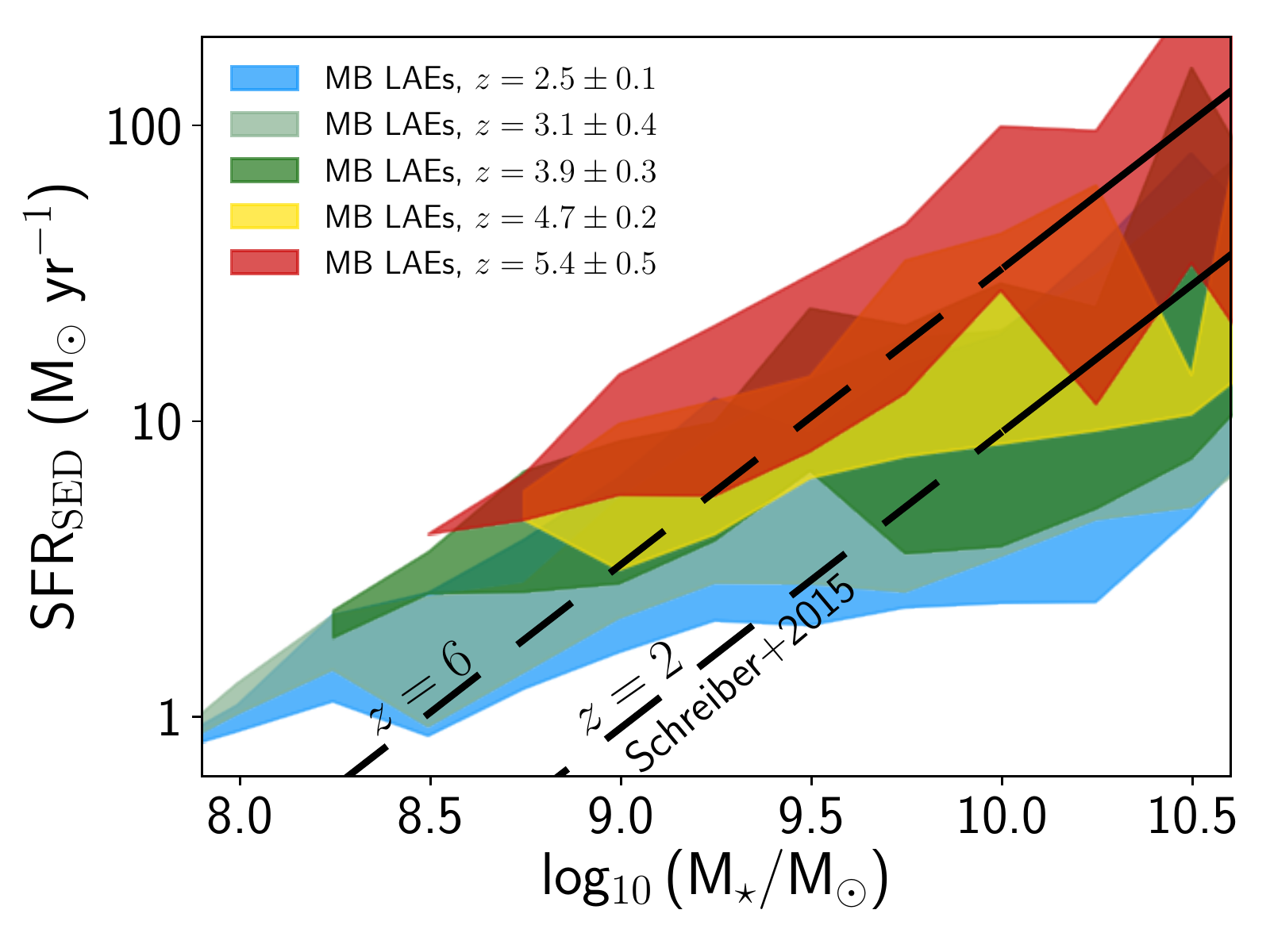} 
\end{tabular}
\caption{{\it Left:} Running average of SFR (derived from Ly$\alpha$ and EW$_0$, see \S\ref{subsec:methods_SFR}) vs M$_\star$ (derived from SED fits, see \S\ref{subsec:methods_SED}). {\it Right:} Same but with SFR derived from {\sc MAGPHYS} (see \S\ref{subsec:methods_SFR}). The SFR-M$_\star$ slopes derived from the two methods are different, with the SED-derived slope being steeper. The difference is likely a consequence of SFR$_{\rm Ly\alpha}$ not being able to reach very low ($<1$\,M$_\odot$ yr$^{-1}$) and very high SFRs ($>20-30$\,M$_\odot$ yr$^{-1}$), but we provide further discussion in \S\ref{sec:SFR_comparison}. For comparison, we show the Main Sequence line for UV-continuum selected sources from \citet{Schreiber2015}, where the dashed lines show the extrapolated values.} \label{fig:contour_sfr_sm19_mstar}
\end{figure*}

\subsection{SFR-M$_\star$ relation and evolution}

We test the dependence of SFR on M$_\star$ in our sample of LAEs and its potential evolution with redshift. In Fig. \ref{fig:sfr_sm19_mstar} we show SFR derived from Ly$\alpha$ and EW$_0$ (see $\S$\ref{subsec:methods_SFR}) vs M$_\star$ (derived from SEDs, \S\ref{subsec:mstar}) for our sample of LAEs and compare with SFRs derived from SED fitting. We compare our measurements with the Main Sequence relation as derived in \cite{Schreiber2015} (converted from Salpeter to Chabrier IMF, extrapolated to low mass ranges when required) and a few studies at different redshifts.

We find that in general there is a relation between SFR and M$_\star$ at all redshifts for LAEs. The relation is relatively shallow when using Ly$\alpha$ SFRs and steeper when using SED SFRs, as can be seen in Fig. \ref{fig:sfr_sm19_mstar}. The relation between SFR and M$_\star$ seems to steepen with increasing redshift for LAEs when using SED SFRs, as can also be seen in Fig. \ref{fig:contour_sfr_sm19_mstar} (right panel). This steepening with increasing redshift also seems to make the SFR-M$_\star$ relation much more in line with the extrapolated relations found for UV-continuum selected sources \citep[e.g.][]{Schreiber2015}.

At $z<4$, we find that LAEs are typically above the Main Sequence relation at their corresponding redshift. This is particularly evident for low stellar masses (M$_{\star}<10^{9.5}$ M$_{\odot}$) although we find that more massive LAEs tend to be within the Main Sequence or even below it, a consequence of the slope of the relation being shallower. At higher redshifts, we find that even at low stellar masses ($10^{9.0-9.5}$ M$_{\odot}$) LAEs are closer to the Main Sequence or that the Main Sequence becomes closer to the relation valid for LAEs, as SFGs may become more LAE-like. Our results therefore suggest that at higher redshifts there is a wider overlap between LAEs and more ``normal" populations of galaxies, as UV-continuum selected galaxies become LAE-like. This could explain the agreement between high-$z$ LAEs and the results of \cite{Salmon2015}. It is nonetheless important to point out (as shown in Fig. \ref{fig:sfr_sm19_mstar}) that the flux limit in Ly$\alpha$ corresponds to a rough cut in SFR and therefore a bias towards higher SFRs at the lowest masses. Similar flux cuts also affect continuum-selected samples, placing them well above the Main Sequence \citep[see e.g.][]{Tasca2015}.

Our results are in good agreement with measurements from Ly$\alpha$-selected samples from \cite{Kusakabe2018} at $z=2$ within error bars. We also compare our results with those presented by \cite{Harikane2018}. While we do not reach such low masses, our results are consistent with LAEs being above the Main Sequence at low stellar masses. With our SC4K sample of LAEs, we can now analyse the evolution of the SFR-M$_\star$ in wide mass ranges at different redshifts, no longer being constrained by single bins or having to stack sources to SED fit the stacked photometry, being able to probe the evolution of the relation within the same sample. 

As previously discussed in \S\ref{sec:SFR_comparison}, there are limitations to different SFR methods, which are important to highlight when comparing the SFR-M$_\star$ relation. SFR$_{\rm Ly\alpha}$ consistently predicts higher SFR than SFR$_{\rm SED}$ for low stellar masses and lower SFR for very high stellar masses. In fact, individual measurements of SFR$_{\rm Ly\alpha}$ seem to fully saturate at $\sim100$\,M$_\odot$\,yr$^{-1}$ with the medians typically not going above $\sim20-30$\,M$_\odot$\,yr$^{-1}$ \citep[see][]{Sobral2018Spectra}. SFR$_{\rm Ly\alpha}$ also implies higher SFRs at lower masses, possibly due to tracing more recent star-formation which would be higher than the one measured from the continuum if LAEs are going through bursts of star-formation. This can be clearly seen with NB LAEs measurements at $z=5.7$, where the low luminosity sample predicts SFRs\,$\approx10$\,M$_\odot$\,yr$^{-1}$. SFR$_{\rm SED}$ may be better suited for such conditions, and as seen in Fig. \ref{fig:sfr_sm19_mstar}, it points towards a relation similar to the \cite{Schreiber2015} extrapolations for the entire mass range we can probe. Nevertheless, we find that the SFRs derived from the two approaches to be consistent, with the same trends being observed from both. In Fig. \ref{fig:contour_sfr_sm19_mstar} we show the running averages for M$_\star$ vs SFR. We find the normalisation of the relation to increase with redshift (left panel) but, as previously discussed, this is mostly driven by detection limits, as we are only capable of reaching down to SFR $<5$\,M$_\odot$ yr$^{-1}$ at $z\sim2$.

\subsection{LAEs: are they ``Main Sequence" galaxies?} \label{sec:MS}

The stellar mass of a galaxy and its star formation rate are correlated in typical galaxies, creating a trend known as the ``Main Sequence"\footnote{Note that galaxies do not evolve along the Main Sequence trend and it is therefore not an evolutionary sequence, see e.g. \cite{MS2019}.} of SFGs \citep[][]{Brinchmann2004,Noeske2007}. A priori, we can naively expect this correlation to occur as the stellar mass of a galaxy is the integral of SFR across time, so the total amount of stars produced will be proportional to the current SFR, assuming a continuous SFR. This dependence can lead to ``tracks" in SED-fitting derived values which lead to a more stringent correlation between SFR and M$_\star$. Galaxies going through periods of intensive star formation, which may be a consequence of bursty star formation, will occupy a region above the Main Sequence. In typical galaxies, SFR and M$_\star$ are in tight correlation and the normalisation of the relation increases with redshift \citep[e.g.][]{Schreiber2015}. Understanding whether the Main Sequence trend holds for LAEs provides important insight into how star formation occurs and how it is driven in this population of predominantly early, primeval galaxies. In principle, we do not expect a Ly$\alpha$-selected sample to span uniformly around the Main Sequence, because we select on emission line strength which at fixed stellar mass always gives high sSFR\,$\equiv$\,SFR/M$_\star$. We therefore do not expect to use LAEs to measure the Main Sequence in an unbiased way, but we can use the comparison to the Main Sequence to determine how LAEs fit in the general galaxy population. Several measurements at $z>2$ have measured the Main Sequence relation by probing M$_{\star}>10^{10}$\,M$_\odot$, with the low mass limit typically rising to M$_{\star}>10^{11}$\,M$_\odot$ at $z>3.5$ \citep[][]{Schreiber2015}, but some recent studies have measured the SFR-M$_\star$ slope and scatter down to M$_\star=10^9$\,M$_\odot$ \citep{Salmon2015}. Our sample of high redshift, typically low M$_\star$ SFGs reaches a region still widely uncharted at these redshift ranges. 

Our results point towards an intensive star formation nature for low mass LAEs at $z<4$, which places them significantly above the extrapolation of the Main Sequence to the lowest masses. A more bursty star-forming nature could explain these SFRs above the Main Sequence. However, we cannot directly infer burstiness from our measurements. More massive LAEs seem to fall within the Main Sequence. At higher redshifts, SFR$_{\rm SED}$-M$_\star$ measurements for LAEs start to resemble more the Main Sequence at all mass ranges. We also find SFR$_{\rm Ly\alpha}$-M$_\star$ to follow a Main Sequence-like relation at $z>4$, except for M$_\star\gtrsim10^{10.5}$, when SFR$_{\rm Ly\alpha}$ seems to saturate, likely due to dust, and is not able to reach SFRs as high as SFR$_{\rm SED}$. This can easily be explained by more massive galaxies showing much higher dust extinction \citep[see e.g.][]{Garn2010, Sobral2012, Whitaker2017}, which at some point might completely absorb Ly$\alpha$ and UV photons in high SFR regions \citep[][]{Sobral2018Spectra}, making it impossible for them to be observed. In such cases, the FIR and some visible and NIR light can still escape, leading to a large discrepancy between SFR$_{\rm SED}$ and SFR$_{\rm Ly\alpha}$. We note that SFR$_{\rm Ly\alpha}$ contains an empirical correction for dust extinction \cite[see][]{SobralMatthee2019}, but this was calibrated for typical LAEs where only moderate to low levels of dust extinction are present leading to Ly$\alpha$ and UV photons being attenuated, but not fully destroyed. At the highest masses, we are likely seeing LAEs with several star-forming regions that may be completely invisible in the UV and Ly$\alpha$ but where at least one region has a hole or a porous ISM \citep[see also][]{Popping2017}.

Overall, we find that the SFR-M$_\star$ relation for LAEs steepens with redshift and that its normalisation also rises with look-back time (see Fig. \ref{fig:contour_sfr_sm19_mstar}). As a consequence, by $z\sim5-6$, LAEs and the general UV-continuum selected population essentially become indistinguishable. This increasing overlap of populations with increasing redshift is also observed in the morphologies and sizes of SFGs, which become LAE-like (compact, $r_e\sim1$\,kpc) towards high redshift \citep{PaulinoAfonso2018} and diverge towards lower redshift as LAEs remain compact at all redshifts. Our results are also fully consistent with the rapid rise of the cosmic average Ly$\alpha$/UV luminosity density ratio with increasing redshift \citep[][]{Sobral2018} which imply that a higher fraction of star-forming galaxies share the properties associated with LAEs, leading to a rise of the cosmic averaged Ly$\alpha$ escape fraction and the cosmic averaged ionisation efficiency, $\xi_{\rm ion}$. Such results are also in agreement with other studies showing a rise of the LAE fraction in UV-selected sources towards $z\sim6$ \citep[][]{CurtisL2012,Schenker2014,Stark2017}, and globally imply that by $z\sim6$ LAEs become representative of the majority of the star-forming population.

\section{Conclusions} \label{sec:conclusions}

In this work, we determined and explored key properties of a large sample of LAEs from the publicly available SC4K survey ($\sim$4000 LAEs at $z\sim2-6$ in the COSMOS field; \citealt{Sobral2018}). We conducted PSF photometry over 34 bands from rest-frame UV to FIR and derived the best-fit SEDs using {\sc MAGPHYS}. We computed SFRs, M$_{\rm UV}$, $\beta$ and M$_{\star}$ for each individual LAE and we provide a full catalogue of SC4K LAEs with all the photometric measurements and derived properties. Our main results are:

\begin{itemize}

\item SC4K LAEs are typically low stellar mass sources (median M$_{\star}$=10$^{9.3^{+0.6}_{-0.5}}$\,M$_\odot$), very blue in the rest-frame UV ($\beta$=-2.1 $^{+0.5}_{-0.4}$) and have low SFRs (SFR$_{\rm Ly\alpha}$: $5.9^{+6.3}_{-2.6}$\,M$_\odot$ yr$^{-1}$;  SFR$_{\rm SED}$: $4.4^{+10.5}_{-2.4}$\,M$_\odot$ yr$^{-1}$).

\item We observe a tight correlation between $\beta$ and M$_{\rm UV}$, qualitatively similar to the one observed in UV-selected samples. The normalisation of this correlation shifts to smaller $\beta$ (bluer) with increasing redshift, which is consistent with a decreasing dust content with increasing redshift in galaxies even for LAEs.

\item Our LAEs are as blue or bluer than UV-selected Lyman Break Galaxies (LBGs) at similar redshifts (up to $\sim$1 dex in the redshift range $z\sim2-6$), suggesting they always constitute the youngest, most metal-poor and/or most dust-poor subset of the UV-selected sources.

\item We find evidence for little to no evolution in the typical Ly$\alpha$ EW$_0$ and the scale parameter $w_0$ with redshift, suggesting the median f$_{\rm esc, Ly\alpha}$ in LAEs is always high and not evolving strongly with redshift.

\item We find that the Ly$\alpha$ $w_0$ (and thus f$_{\rm esc, Ly\alpha}$) for LAEs declines with increasing stellar mass, implying that f$_{\rm esc, Ly\alpha}$ is highest for the lowest stellar mass LAEs and lowest for the most massive LAEs. A similar trend is found with rest-frame UV luminosity, where the faintest LAEs have the highest typical EWs and the highest f$_{\rm esc, Ly\alpha}$.

\item We explore extreme EW$_0$ measurements in our large sample of LAEs and find 45 non-AGN LAEs with EW$_0>240$\,{\AA} at a 3\,$\sigma$ level, resulting in a number density $(7\pm1)\times10^{-7}$\,Mpc$^{-3}$. These extreme emitters are incredibly rare but can provide insight into extreme Ly$\alpha$ emission that is neither purely from typical star-formation or AGN.

\item By using Ly$\alpha$ EW$_0$ to infer f$_{\rm esc, Ly\alpha}$ \citep[][]{SobralMatthee2019} we compute Ly$\alpha$ SFRs which are independent of SED fitting measurements and we compare both. Ly$\alpha$ and SED-fitting based SFRs show a remarkable agreement for M$_\star=10^{9-10}$\,M$_\odot$ and SFR$_{\rm SED}=1-10$\,M$_\odot$ yr$^{-1}$. SFR$_{\rm Ly\alpha}$ predicts lower SFRs at more massive regimes, likely due to not being sensitive to heavily obscured parts of very massive galaxies.

\item LAEs show a relation between stellar mass and SFR at all redshifts, but this is typically shallower than the relation found for the general star-forming population. We also find that the relation steepens and rises with increasing redshift for LAEs.

\item LAEs are typically above the ``Main Sequence" at $z<4$ and M$_{\star}$<10$^{9.5}$\,M$_\odot$, indicating LAEs are experiencing more intense star formation than the general population of galaxies of similar mass at similar redshifts, with one possible explanation being a bursty star-formation nature of LAEs. For higher masses and redshifts, this offset decreases, implying a larger overlap between LAEs and more ``normal" SFGs.
\end{itemize}

Overall, we find that LAEs are typically very young, low mass galaxies, albeit they still span an important range of properties, and within the LAE population there are important trends with stellar mass and UV luminosity. Typical properties of LAEs seem to have little evolution between $z=2$ and $z=6$, although they still become bluer and the relation between SFR and stellar mass steepens and rises slightly. By $z\gtrsim4$, the overlap between LAEs and the more general UV-selected population becomes significant and by $z\sim6$ they seem to become undistinguishable, as typical SFGs essentially become LAE-like. Our results reveal how galaxies selected as LAEs constitute mostly the youngest, most primeval galaxies at any redshift, and also that LAEs are ideal sources to study the dominant population of SFGs towards $z\gtrsim6$ and therefore also likely the population that re-ionised the Universe.

\section*{Acknowledgements}

We thank the anonymous referee for the valuable feedback that significantly improved the quality and clarity of this paper. SS and JC acknowledge studentships from Lancaster University. APA acknowledges support from Funda\c{c}\~{a}o para a Ci\^{e}ncia e a Tecnologia through the project PTDC/FIS-AST/31546/2017. The authors would like to thank Ali Khostovan, Sara Perez Sanchez, Alex Bennett and Tom Rose for contributions and discussions in the early stages of this work. Based on data products from observations made with ESO Telescopes at the La Silla Paranal Observatory under ESO programme ID 179.A-2005 and on data products produced by CALET and the Cambridge Astronomy Survey Unit on behalf of the UltraVISTA consortium.

Finally, the authors acknowledge the unique value of the publicly available analysis software {\sc TOPCAT} \citep{Taylor2005} and publicly available programming language {\sc Python}, including the {\sc numpy}, {\sc pyfits}, {\sc matplotlib}, {\sc scipy} and {\sc astropy} \citep{Astropy2013} packages. This work is based on the public SC4K sample of LAEs \citep[][]{Sobral2018} and we release the \href{https://goo.gl/q9yfKo}{full catalogue with all the photometry and properties derived in this paper}, in electronic format, along with the relevant tables.

\bibliographystyle{mnras}
\bibliography{myBib}


\appendix

\section{The full SC4K catalogue with PSF photometry and all derived quantities} \label{ap:cat}

We provide the \href{https://goo.gl/q9yfKo}{full catalogue of SC4K LAEs in electronic format ({\sc fits} format)} with PSF photometry and photometric errors in all bands, along with all the properties obtained in this paper.

\section{Additional plots and tables}

\begin{figure*}
  \centering
  \includegraphics[width=\textwidth]{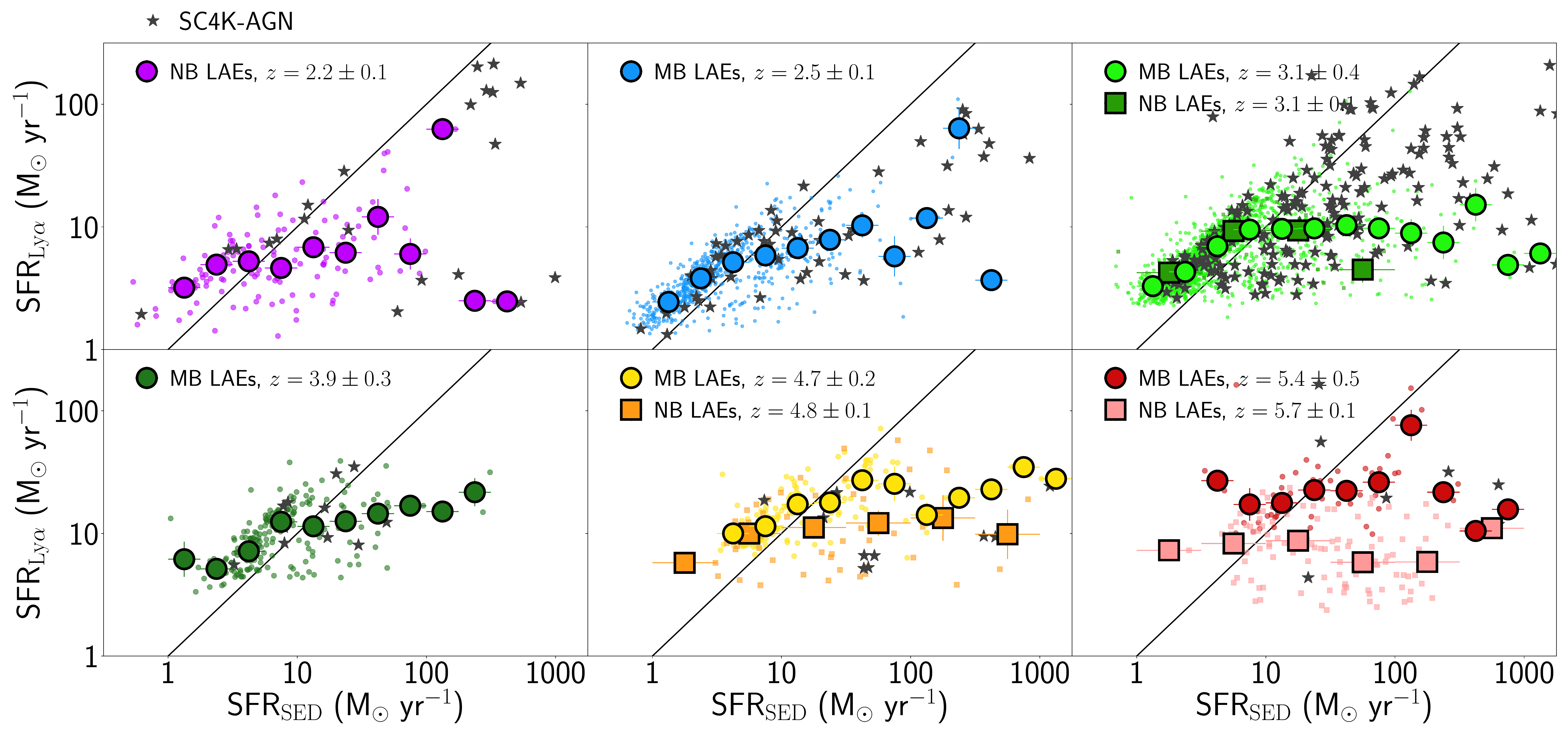}
  \caption{Emission line-based SFR vs SED-fitting SFR for the full sample of LAEs at different redshift ranges. Coloured circles (squares) are the median bin for MBs (NBs) and individual points are plotted as scatter in the background. The black line is the 1-to-1 ratio. While the two approaches roughly follow the 1-to-1 ratio, there are some key differences. Similar to what is observed in Fig. \ref{fig:SFR_SM19vsSFR_SED_Mstar}, median SFR$_{\rm Ly\alpha}$ is slightly higher than SFR$_{\rm SED}$ for SFR$_{\rm SED}<10$\,M$_\odot$ yr$^{-1}$. However, SFR$_{\rm Ly\alpha}$ seems to saturate at median SFR$_{\rm Ly\alpha}\approx10-30$\,M$_\odot$ yr$^{-1}$.}
  \label{fig:SFR_SM19vsSFR_SED}
\end{figure*}

\begin{figure}
  \centering
  \includegraphics[width=0.5\textwidth]{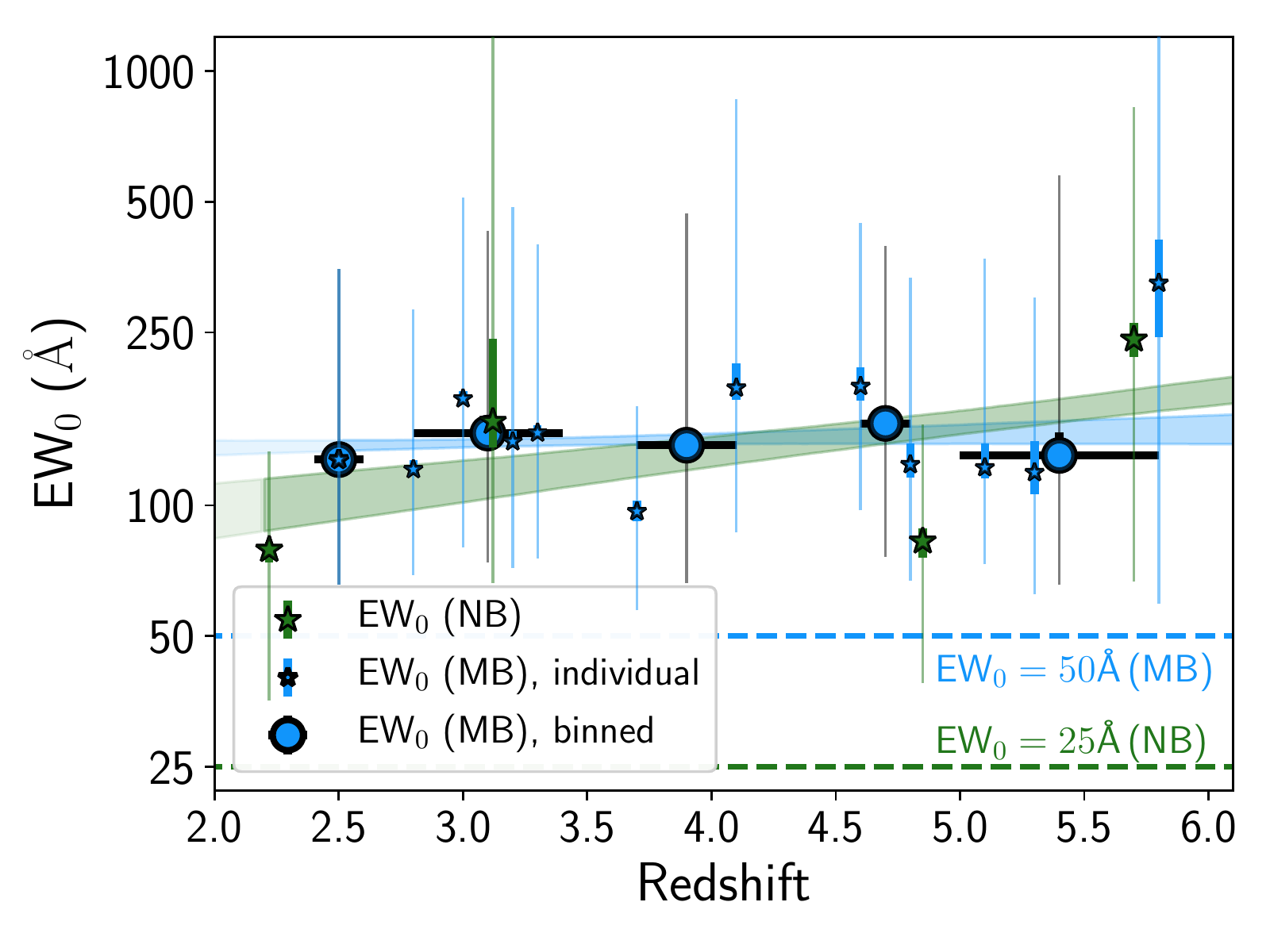}
  \caption{Global median EW$_{0}$ evolution with redshift. The median EW$_{0}$ values for medium (narrow) bands are shown as blue circles (green stars). Blue stars are the measurements for individual MBs. The thin (thick) error bars are the 16th and 84th percentiles of the EW$_{0}$ distribution (divided by the Poissonian error $\sqrt{N}$). The median and errors of EW$_0$ can be found in Table \ref{tab:overview}. Blue (green) shaded region is the 1\,$\sigma$ contour obtained by perturbing the EW$_{0}$ within the thick error bars for medium (narrow) band selected LAE. We find evidence of little EW$_{0}$ evolution with redshift for the global sample of LAEs, with the median EW$_{0}$ remaining roughly constant at $\sim140$\,\AA, although there is a tentative higher EW$_0$ at $z=5.7$, albeit with large error bars.}
  \label{fig:EW_evolution}
\end{figure}

In Fig. \ref{fig:SFR_SM19vsSFR_SED} we show SFR$_{\rm Ly\alpha}$ vs SFR$_{\rm SED}$ in 6 independent redshift intervals (see \S\ref{sec:SFR_comparison} for discussion). In Fig. \ref{fig:EW_evolution} we show the evolution of median EW$_0$ with redshift. We provide the full measurements of $w_0$ for different ranges of redshifts and galaxy properties (M$_\star$ and M$_{\rm UV}$) in Table \ref{tab:w0}. 

%
%

%
%

\begin{table*}
\begin{center}
\caption{EW$_0$ scale length ($w_0$) for different redshift bins, derived as fully detailed in \S\ref{sec:w0}. (1) Subset of SC4K; (2) Filter type (MB/NB) and whether this measurement is for LAEs selected from an individual filter or from multiple filters binned together; (3) $w_0$ derived directly from observed counts; (4) $\chi^2_{\rm red}$ of (3); (5) $w_0$ derived directly from observed counts [EW$_0<240$\,\AA]; (6) $\chi^2_{\rm red}$ of (5); (7) $w_0$ derived by perturbing EW$_0$; (8) $\chi^2_{\rm red}$ of (7); (9) $w_0$ derived by perturbing EW$_0$ [EW$_0<240$\,\AA]; (10) $\chi^2_{\rm red}$ of (9); $^\star$Not enough sources to constrain $w_0$ (less than 3 bins with 5 sources). $^\dagger$Fit done for EW$_0>100$\,\AA\, as discussed in \S\ref{sec:w0} as we are significantly incomplete for low EW$_0$, so we only fit exponential decay after the distribution peak at $\sim100$\,\AA.} \label{tab:w0}
\begin{tabular}{c | ccccccccc}
\hline
(1)& (2) & (3) & (4) & (5) & (6) & (7) & (8) & (9) & (10)\\
\multicolumn{1}{c|}{Subset} &
\multicolumn{1}{c|}{Filters} &
\multicolumn{1}{c|}{$w_0$} &
\multicolumn{1}{c|}{$\chi^2_{\rm red}$} &
\multicolumn{1}{c|}{$w_{\rm 0, [EW<240\text{\normalfont\AA}]}$} &
\multicolumn{1}{c|}{$\chi^2_{\rm red}$}&
\multicolumn{1}{c|}{$w_{\rm 0,p}$} &
\multicolumn{1}{c|}{$\chi^2_{\rm red}$} &
\multicolumn{1}{c|}{$w_{\rm 0,p, [EW<240\text{\normalfont\AA}]}$} &
\multicolumn{1}{c|}{$\chi^2_{\rm red}$}\\
&  & (\AA) &  & (\AA) &  & (\AA) &  & (\AA) & \\
\hline
$z=2.5\pm0.1$ & MB, single & $104^{+13}_{-13}$ & 2.31 & $79^{+15}_{-15}$ & 2.77 & $117^{+12}_{-12}$ & 0.73 & $94^{+12}_{-12}$ & 0.12 \\
$z=2.8\pm0.1$ & MB, single & $98^{+12}_{-12}$ & 0.59 & $88^{+12}_{-12}$ & 0.37 & $108^{+12}_{-13}$ & 0.15 & $100^{+17}_{-15}$ & 0.28 \\
$z=3.0\pm0.1$ & MB, single & $172^{+14}_{-14}$ & 1.04 & $120^{+22}_{-22}$ & 1.74 & $170^{+16}_{-16}$ & 0.72 & $126^{+22}_{-17}$ & 0.37 \\
$z=3.2\pm0.1$ & MB, single & $109^{+14}_{-14}$ & 1.77 & $97^{+16}_{-16}$ & 1.17 & $128^{+15}_{-14}$ & 0.34 & $111^{+20}_{-18}$ & 0.18 \\
$z=3.3\pm0.1$ & MB, single & $113^{+13}_{-13}$ & 1.76 & $118^{+19}_{-19}$ & 1.21 & $139^{+13}_{-14}$ & 0.40 & $113^{+16}_{-14}$ & 0.42 \\
$z=3.7\pm0.1$ & MB, single & $83^{+24}_{-24}$ & 1.64 & $72^{+26}_{-26}$ & 1.93 & $90^{+24}_{-19}$ & 0.20 & $80^{+24}_{-17}$ & 0.12 \\
$z=4.1\pm0.1$ & MB, single & $257^{+41}_{-41}$ & 0.66 & $216^{+100}_{-100}$ & 0.91 & $251^{+81}_{-65}$ & 0.31 & $171^{+149}_{-54}$ & 0.08 \\
$z=4.6\pm0.1$ & MB, single & $486^{+183}_{-183}$ & 0.65 & $318^{+413}_{-413}$ & 1.30 & $600^{+325}_{-223}$ & 0.44 & $413^{+43491}_{-232}$ & 0.08 \\
$z=4.8\pm0.1$ & MB, single & $93^{+17}_{-17}$ & 0.42 & $82^{+14}_{-14}$ & 0.18 & $119^{+41}_{-31}$ & 0.21 & $94^{+30}_{-22}$ & 0.12 \\
$z=5.0\pm0.1$ & MB, single & $108^{+26}_{-26}$ & 0.75 & $85^{+24}_{-24}$ & 0.75 & $143^{+84}_{-42}$ & 0.40 & $103^{+39}_{-24}$ & 0.19 \\
$z=5.3\pm0.1$$^\star$ & MB, single & - & - & - & - & - & - & - & - \\
$z=5.8\pm0.1$$^\star$ & MB, single & - & - & - & - & - & - & - & - \\
\hline
$z=2.2\pm0.1$ & NB, single & $174^{+95}_{-95}$ & 4.95 & $174^{+95}_{-95}$ & 4.95 & $143^{+22}_{-19}$ & 0.52 & $131^{+26}_{-19}$ & 0.62 \\
$z=3.1\pm0.1$$^\star$ & NB, single & - & - & - & - & - & - & - & - \\
$z=4.8\pm0.1$ & NB, single & $86^{+24}_{-24}$ & 1.05 & $86^{+24}_{-24}$ & 1.05 & $151^{+98}_{-50}$ & 0.48 & $101^{+31}_{-26}$ & 0.32 \\
$z=5.7\pm0.1$ & NB, single & $355^{+71}_{-71}$ & 0.93 & $188^{+57}_{-57}$ & 0.70 & $477^{+154}_{-106}$ & 0.45 & $124^{+34}_{-27}$ & 0.14 \\
\hline
$z=2.5\pm0.1$ & MB, bin & $104^{+13}_{-13}$ & 2.31 & $79^{+15}_{-15}$ & 2.77 & $117^{+12}_{-12}$ & 0.73 & $94^{+12}_{-12}$ & 0.12 \\
$z=3.1\pm0.4$ & MB, bin & $134^{+11}_{-11}$ & 2.00 & $109^{+12}_{-12}$ & 0.85 & $149^{+11}_{-11}$ & 2.26 & $116^{+13}_{-12}$ & 1.04 \\
$z=3.9\pm0.3$ & MB, bin & $118^{+18}_{-18}$ & 1.29 & $90^{+21}_{-21}$ & 1.69 & $120^{+18}_{-17}$ & 0.20 & $103^{+21}_{-18}$ & 0.15 \\
$z=4.7\pm0.2$ & MB, bin & $119^{+21}_{-21}$ & 1.17 & $93^{+19}_{-19}$ & 0.83 & $158^{+34}_{-27}$ & 0.18 & $114^{+43}_{-23}$ & 0.14 \\
$z=2.5\pm0.1$ & MB, bin & $95^{+24}_{-24}$ & 1.66 & $70^{+21}_{-21}$ & 1.97 & $125^{+40}_{-31}$ & 0.24 & $90^{+22}_{-18}$ & 0.21 \\
\hline
Full sample & MB, bin & $130^{+11}_{-11}$ & 3.57 & $100^{+11}_{-11}$ & 1.55 & $143^{+10}_{-11}$ & 4.01 & $110^{+11}_{-11}$ & 0.99 \\
Full sample & NB, bin & $109^{+13}_{-13}$ & 1.53 & $102^{+14}_{-14}$ & 0.98 & $151^{+18}_{-17}$ & 0.45 & $102^{+15}_{-14}$ & 0.15 \\
Full sample & MB+NB, bin & $129^{+11}_{-11}$ & 4.19 & $99^{+11}_{-11}$ & 1.22 & $147^{+11}_{-11}$ & 4.46 & $109^{+11}_{-11}$ & 0.96 \\
\hline
8<$\log_{10}\,$(M$_{\star}$/M$_{\odot}$<9$^\dagger$ & MB+NB, bin & $175^{+14}_{-14}$ & 1.62 & $179^{+84}_{-84}$ & 2.77 & $264^{+14}_{-16}$ & 0.55 & $530^{+582}_{-163}$ & 1.42 \\
9<$\log_{10}\,$(M$_{\star}$/M$_{\odot}$<10 & MB+NB, bin & $85^{+11}_{-11}$ & 2.87 & $74^{+11}_{-11}$ & 1.75 & $101^{+11}_{-10}$ & 2.03 & $89^{+11}_{-11}$ & 0.96 \\
10<$\log_{10}\,$(M$_{\star}$/M$_{\odot}$<11 & MB+NB, bin & $77^{+13}_{-13}$ & 2.14 & $60^{+13}_{-13}$ & 2.32 & $89^{+13}_{-13}$ & 0.81 & $68^{+12}_{-11}$ & 0.66 \\
\hline
-20<M$_{\rm UV}$<-19$^\dagger$ & MB+NB, bin & $182^{+13}_{-13}$ & 1.35 & $253^{+103}_{-103}$ & 1.29 & $263^{+15}_{-15}$ & 0.50 & $470^{+212}_{-126}$ & 0.97 \\
-21<M$_{\rm UV}$<-20 & MB+NB, bin & $77^{+10}_{-10}$ & 1.52 & $71^{+11}_{-11}$ & 1.00 & $93^{+10}_{-10}$ & 1.09 & $90^{+11}_{-11}$ & 1.27 \\
-22<M$_{\rm UV}$<-21 & MB+NB, bin & $55^{+11}_{-11}$ & 1.78 & $50^{+11}_{-11}$ & 1.91 & $63^{+11}_{-11}$ & 0.39 & $58^{+11}_{-10}$ & 0.07 \\

\hline
\end{tabular}
\end{center}
\end{table*}


\bsp	
\label{lastpage}
\end{document}